\documentclass[iop]{emulateapj}
\usepackage{hyperref}
\usepackage{graphicx}
\usepackage{longtable}
\usepackage{amsmath}


\setkeys{Gin}{draft=false}

\shortauthors{{Winter and Ledbetter}}
\shorttitle{{Solar Radio Bursts and SEP Properties}}

\begin{document}



\title{Type II and Type III Radio Bursts and their \\Correlation with Solar Energetic Proton Events}  

\author{L.M. Winter}
\affil{Atmospheric and Environmental Research, Superior, CO, USA.}
\email{lwinter@aer.com}

\and
\author{K. Ledbetter}
\affil{Wellesley College, Wellesley, MA, USA.}

\begin{abstract}

Using the Wind/WAVES radio observations from 2010--2013, we present an analysis of the 123 decametric-hectometric (DH) type II solar radio bursts during this period, the associated type III burst properties, and their correlation with solar energetic proton (SEP) properties determined from analysis of the Geostationary Operational Environmental Satellite (GOES) observations.   We present a useful catalog of the type II burst, type III burst, Langmuir wave, and proton flux properties for these 123 events, which we employ to develop a statistical relationship between the radio properties and peak proton flux that can be used to forecast SEP events. We find that all SEP events with a peak $> 10$\,MeV flux above 15\,protons\,cm$^{-2}$\,s$^{-1}$\,sr$^{-1}$ are associated with a type II burst and virtually all SEP events, 92\%, are also associated with a type III radio burst.  Based on a principal component analysis, the radio burst properties that are most highly correlated with the occurrence of gradual SEP events and account for the most variance in the radio properties are the type III burst intensity and duration.  Further, a logistic regression analysis with the radio-derived principal component (dominated by the type III and type II radio burst intensity and type III duration) obtains SEP predictions approaching the human forecaster rates, with a false alarm rate of 22\%,  a probability of detection of 62\%, and with 85\% of the classifications correct.  Therefore, type III radio bursts that occur along with a DH type II burst are shown to be an important diagnostic that can be used to forecast SEP events.

\end{abstract}

\section{Introduction}
Significant increases in the solar energetic particle flux, associated with solar flares and coronal mass ejections, can cause major disruptions to human technology and pose health risks to astronauts, as well as passengers and crew on polar flights.  Increased prediction accuracy and warning time are crucial to mitigate the effects of solar energetic proton (SEP) events.  For example, accurate SEP predictions allow satellite operators the possibility of shutting down critical satellite systems, preventing failure modes, and airlines the possibility of re-routing commercial polar flights. 

The most impactful SEP events, those with the longest duration and highest particle fluence, are believed to be created from shocks associated with coronal mass ejections (CMEs, e.g.,  \citealt{:fk}). CME-driven shocks are connected with the production of type II radio bursts (e.g.,
\citealt{1985srph.book..333N}), whereas type III radio bursts are associated with solar flares \citep{1950AuSRA...3..541W}.  Thus, prediction models utilizing the occurrence or fluence of radio bursts (e.g., the NOAA SWPC model, presented in
\citealt{Balch1999}
and \citealt{2008SpWea...601001B},
 and the \citealt{2009SpWea...704008L} model) tend to achieve higher probability of detection rates than alternate methods, particularly for events that are magnetically well-connected to the Earth \citep{2011SpWea...907003N}.  The focus of this research is to use the available observational data of solar radio bursts and SEP events  to establish connections between observable solar radio phenomena  and solar energetic particle events.

Our immediate focus is an analysis of the space based solar radio data, measuring the properties of decametric-hectometric (DH) type II radio bursts, type III radio bursts, and local Langmuir waves.  While much work has been done linking the properties of type II radio bursts and coronal mass ejections (e.g., \citealt{2001JGR...10629219G}), less is known about how the low frequency type II radio burst properties relate to energetic proton events and low frequency type III bursts.  However, it has been established that large gradual SEP events are associated with the DH type II bursts \citep{2002ApJ...572L.103G, 2004ApJ...605..902C}.
There is also a known correlation in the association of decametric-hectometric type III bursts \citep{2009ApJ...690..598C} that warrants further investigation.  

Large SEP events are known to be associated with complex type III bursts accompanying type II bursts \citep{2009ApJ...690..598C} and while there has previously been no use of DH type II bursts in SEP forecasting, type III bursts are used in the \citet{2009SpWea...704008L} model through the 1 MHz radio flux.  Finally, Langmuir waves are produced from electrons passing by the satellite and oscillating at the local plasma frequency.  The same electrons that cause the detection of intense Langmuir waves by the satellite are also believed to have been accelerated at the flare site where the type III burst is produced (e.g.,\citealt{1978ApJ...223..605N}).  

In this paper, we identify statistical correlations between the low frequency interplanetary type II radio burst properties (e.g., fluence and frequency drift rates, which are related to  shock speed, e.g., \citealt{1998JGR...10329651R}), type III burst properties (e.g., intensity and duration), Langmuir waves (i.e., intensity), 
and the proton event properties (e.g., energy spectra, peak intensities, duration).  We describe the sample  of type II bursts from Jan 2010- May 2013 and data analyzed in \S~\ref{data}.  Our data analysis is described in \S~\ref{data analysis}.  In \S~\ref{properties}, we present the radio and proton flux characteristics of the sample.   In \S~\ref{pca}, 
 we conduct a principal component analysis to create a radio index for use in forecasting SEP events. We then perform a logistic regression analysis between the radio index and the peak proton flux level to distinguish between SEP and non-SEP events. Our results reveal a relationship between the radio { index} and the peak proton flux level that predicts with good accuracy whether the peak proton level will exceed the NOAA SWPC warning level. We discuss how our results can be used for forecasting SEP events in \S~\ref{forecast}, along with the limitations of our method.  Finally, we present our conclusions in \S~\ref{conclusions}.


\begin{figure}
\begin{center}
\includegraphics[width=0.95\linewidth]{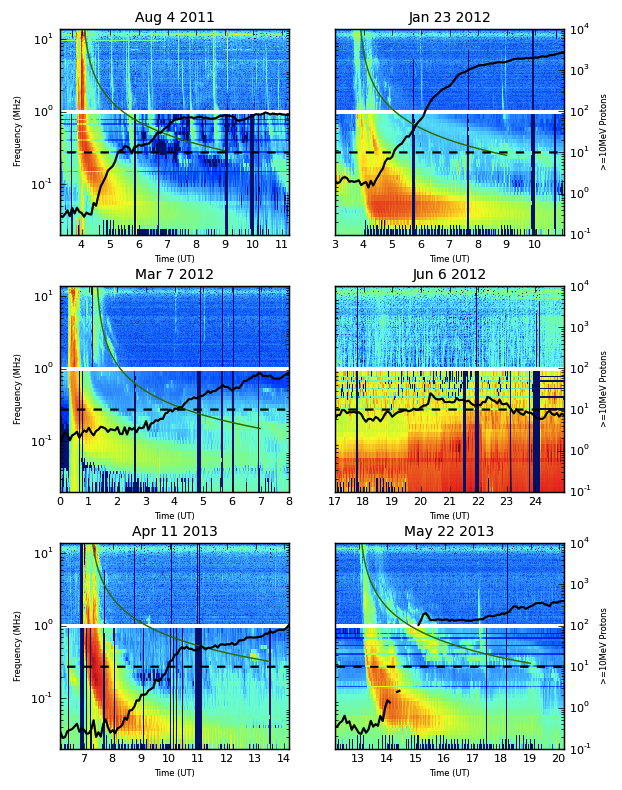}
\caption{Examples of the WIND/WAVES dynamic spectra during SEP events.  The black lines show the $> 10$\,MeV GOES proton flux in pfu superimposed.  A green line is used to easily distinguish the type II solar radio burst in the data.  The dashed line shows the $> 10$\,MeV 10\,pfu threshold of the NOAA SWPC warning level for SEP events. { The white line indicates frequencies where there is a gap in the WIND/WAVES data coverage.}}\label{fig-radioseps}
\end{center}
\end{figure}

\begin{figure}
\begin{center}
\includegraphics[width=0.95\linewidth]{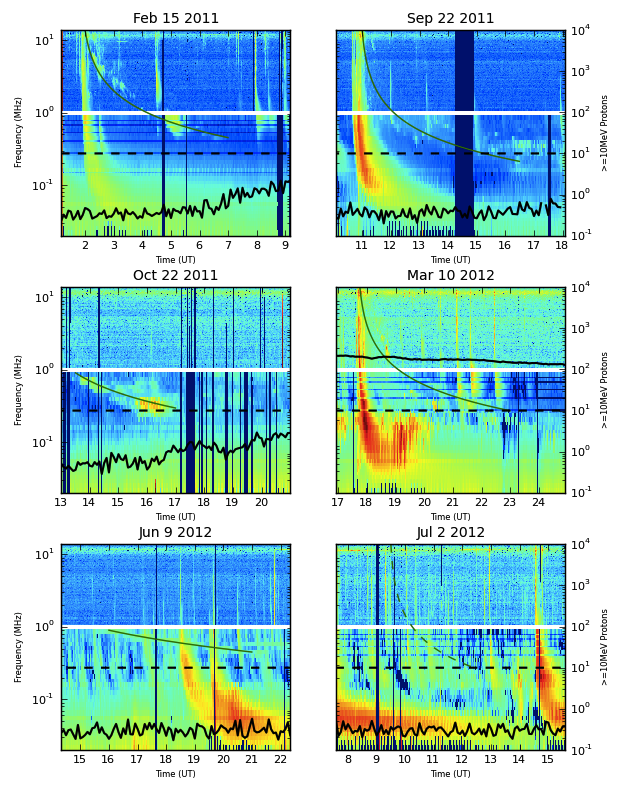}
\caption{Examples of the WIND/WAVES dynamic spectra during type II radio bursts.  For all but the March 10th, 2012 event (where the burst occurs after an SEP event is already underway), an SEP event is not associated with the type II burst.  The figure description is the same as for Figure~\ref{fig-radioseps}.}\label{fig-radiononseps}
\end{center}
\end{figure}

\section{Observations}\label{data}
Our sample includes all type II radio bursts identified in the {\it Type II and IV burst lists} available from the NASA Wind/WAVES website.  The solar radio burst list includes all possible type II and type IV bursts detected by Wind/WAVES and Solar Terrestrial Relations Observatory (STEREO)/WAVES instruments.  Our study includes an analysis of the type II bursts from Jan 2010--May 2013, 123 bursts in total.  The selected period includes overlapping mission time from the Wind and STEREO missions.  While STEREO's archives include radio data starting in 2007, only 5 bursts are detected from 2007-2010, none of which are associated with SEP events.  Therefore, they are not included in the study of SEP/radio bursts.  In this paper, we include an analysis of the Wind/WAVES data but in future work we will present a comparison with the STEREO/WAVES data to evaluate their use in developing a real-time SEP forecaster based on type II and III radio burst properties.      

For the sample of 123 type II bursts, we analyzed the Wind/WAVES dynamic spectra.  The Wind/WAVES instrument \citep{1995SSRv...71..231B} includes three detectors, RAD1 (20--1040 kHz), RAD2 (1.075--13.825 MHz), and the thermal noise receiver (TNR; 4-245 kHz).  The calibrated one-minute averages from each of these detectors were downloaded for the duration of each of the 123 bursts from the Wind/WAVES data archive.  The data in these files are recorded as the ratio (R) to the background (B) values in units of $\mu$V\,Hz$^{-1/2}$ and are converted to solar flux units (1 sfu $= 10^{-22}$\,W\,m$^{-2}$\,Hz$^{-1}$) as: \begin{math} {\rm I (sfu)} =  10^{10} (\rm{R} \times \rm{B})^2 / {Z_0 \times A} \end{math}, where Z$_0$ is the impedance of free space (377 Ohms), A is the area in m$^2$ of the antenna (e.g., RAD1 has an area of 1225 m$^2$), R is the ratio in the IDL save file, and B is the background value in the IDL save file (details included in \citealt{2010JGRA..115.6102H}).  These data provide the basis of our analysis, described in \S~\ref{radio analysis}.    

The sample of type II radio bursts is compared with the NOAA SWPC {\it Solar Proton Events Affecting the Earth Environment} list of SEP events.  SEP events in this catalog are defined as having $> 10$\,proton flux units (pfu = protons\,cm$^{-2}$\,s$^{-1}$\,sr$^{-1}$) integral flux at energies $> 10$\,MeV, in the GOES detectors.  While alternative methods and catalogs are available for SEP events, for example through ESA's Solar Energetic Particle Environment Modeling (SEPEM, is available at: {http://dev.sepem.oma.be}.
), we choose the NOAA SWPC criteria in order to be consistent when comparing to the well-established NOAA warning levels. 

  To fully characterize the SEP properties associated with the type II radio bursts, we include an analysis of the proton flux properties from the GOES satellites.  From Jan 2010 -- May 2013, the proton measurements were from the Space Environment Monitor (SEM) on GOES 13, 14, and 15. We use the integral proton flux measurements, which are available as 5-min averages in  $> 1$\,MeV, $> 5$\,MeV, $> 10$\,MeV, $> 30$\,MeV, $> 50$\,MeV, $> 60$\,MeV, and $> 100$\,MeV.  The data were downloaded from the GOES Space Environment Monitor Data Access website at NOAA's NGDC.  Analysis of these data are included in \S~\ref{goes analysis}.

\section{Data Analysis}\label{data analysis}
\subsection{Wind/WAVES Radio Data}\label{radio analysis}
The processed RAD1, RAD2, and TNR data from Wind/WAVES were downloaded and used to construct dynamic spectra for time periods encompassing each of the 123 type II radio bursts occurring from Jan 2010--May 2013.  Analyses of the type II and type III properties utilized the RAD1 and RAD2 data, while the TNR data were primarily used to determine the Langmuir wave properties.
Properties of the bursts are listed in Table~\ref{table-radiointensity}.  The type III bursts associated
with type II bursts, as listed in Table~\ref{table-radiointensity}, are defined as the strongest type III burst within 2 hours before the start of the type II burst.  In Figures~\ref{fig-radioseps} and \ref{fig-radiononseps} example dynamic radio spectra are shown for type II bursts with and without associated SEP events.  In many cases, a strong type III burst is seen before a rise in proton flux.  In several cases where the proton flux did not meet the NOAA SWPC $> 10$\,MeV threshold for an SEP event, it does meet alternative criteria such as those used in  \citet{2014SoPh..289.3059R} for $> 25$\,MeV SEP events ({ with an event defined with a proton flux $> 10^{-4}$\,(MeV\,s\,cm$^{-2}$\,sr)$^{-1}$}).

Following \citet{2010ApJ...710L..58L}, who present a method for automated analyses of type II bursts tested on Learmoth Solar Radio Obseratory data, the data are transformed into 1/frequency space.  Using this transformation, the type II and type III bursts are linear in time.  Further processing included removing gaps from the detector, boxcar smoothing, and increasing the contrast in the image with histogram equalization.  Following this, the properties of the type II and type III bursts were determined by finding the local intensity maxima at each frequency.  Linear functions were then fit to the local maxima of the strongest type III and type II bursts present.  For the type III bursts, in addition to the slope, we measured the integrated intensity of the burst by integrating along the fitted line between the points where the flux falls to 15\% of the logarithm of peak intensity (the flux at the highest local maximum).  Additionally, we measure the duration of the type III burst at the time where the 1\,MHz signal exceeds 6\,dB or four times the background level \citep{2003GeoRL..30.8018M}.

For the type II properties, the integral intensity of the bursts was determined in the same way as for the type III bursts (integrated flux within 15\% of the highest local peak intensity).  The frequency range, peak intensity (the highest flux local maximum), duration of the burst, and integrated intensity were measured for the type II bursts. Also, since the frequency drift rate of type II bursts is related to the shock speed, we also calculated the drift rate through the type II starting frequency.

Starting frequencies for the type II bursts were taken from the {\it Type II and IV burst lists}. The frequencies are given with uncertainty of $\pm 1$\,MHz. However, these observations are limited by the fact that many bursts start above the range of Wind's radio receivers. The maximum starting frequency that can be observed is 16 MHz for STEREO/WAVES and 14 MHz for Wind/WAVES; therefore, all points at 16 MHz (or 14 MHz if observed by Wind only) must be considered to indicate a starting frequency $\ge 16$ MHz (14 MHz).

Visual inspection of the spectra showed that fifteen of the bursts did not appear to have the starting frequency given in the WAVES burst catalog. This was due to either the burst appearing at a higher frequency than tabulated on one or more receivers (WIND, STEREO A or STEREO B) or the burst appearing at a lower frequency than tabulated due to the appearance of harmonics above the actual burst frequency (e.g., Feb 15, 2011). These fifteen points were not included in further analyses of drift rate and are marked with a `?' in Table~\ref{table-radiointensity}.

To determine the type II burst frequency drift rate, we use the relationship between drift rate and starting frequency derived by \citet{2005ESASP.592..393A}. For the RAD2 receiver on WIND, they find: \begin{math}-{\rm df}/{\rm dt} = 5.50 \times 10^{-5} {\rm f}_s^{1.28} {\rm MHz}\,{\rm s}^{-1}\end{math}. To confirm that this equation gives a good approximation of drift rate for our data set, we compared the drift rate predicted by the equation to the drift rate obtained from the radio spectra, using the same procedure as outlined in \citet{2005ESASP.592..393A}, for a representative sample of five bursts with starting frequencies ranging from 2-16\,MHz. In this procedure, Gaussian functions are fit to the flux density profile of each burst in 60\,s time bins.  Drift rate is found by linear regression fits to the central frequency versus time from the bins.  We found that our calculations of drift rate matched the prediction  within 1-30\%, with higher drift rates measured for bursts with starting frequencies above 14\,MHz. Therefore, we used the drift rate calculation (with results discussed in \S~\ref{radio-statistics}) from Agilar-RodriguezÕ equation, but note that bursts with a 16 MHz starting frequency could have a higher drift rate by factors of 30\%.

The final parameters measured in the radio data were those of the Langmuir waves.  The detection of local Langmuir waves in the Wind TNR data is a good indicator that the active site accelerating the electrons, and presumably the SEPs, is magnetically well-connected with the Earth. We can also assume a good connection to the CME-driven shock, because flare sites are commonly observed under the center of the CME span \citep{2009IAUS..257..233Y}.  To characterize the properties of the Langmuir waves, we determined the peak intensity from 9-49\,kHz in the TNR data.  This frequency band also exhibits contributions to the flux from type III bursts in some cases.  The intensity of peaks in the time derivative, caused by Langmuir waves, were characterized, filtering out the additional flux from the type III bursts.

\begin{figure}
\begin{center}
\includegraphics[width=0.45\linewidth]{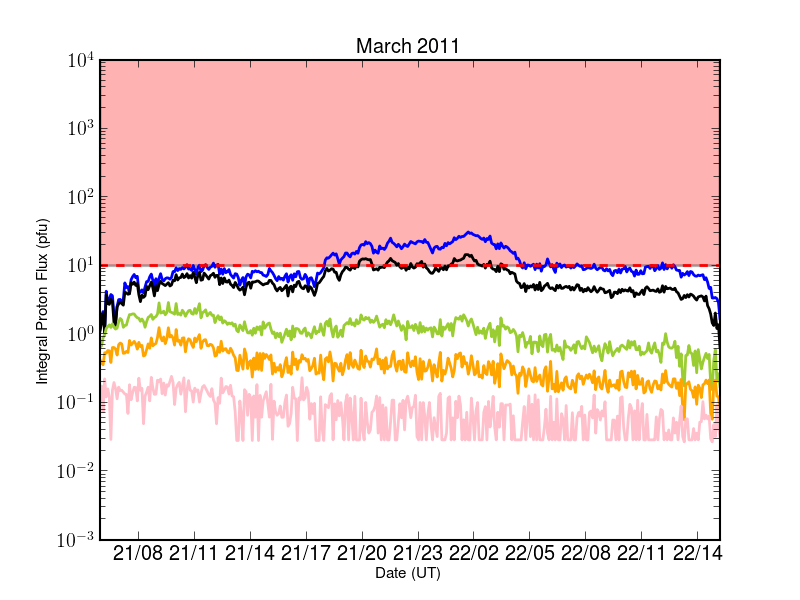}
\includegraphics[width=0.45\linewidth]{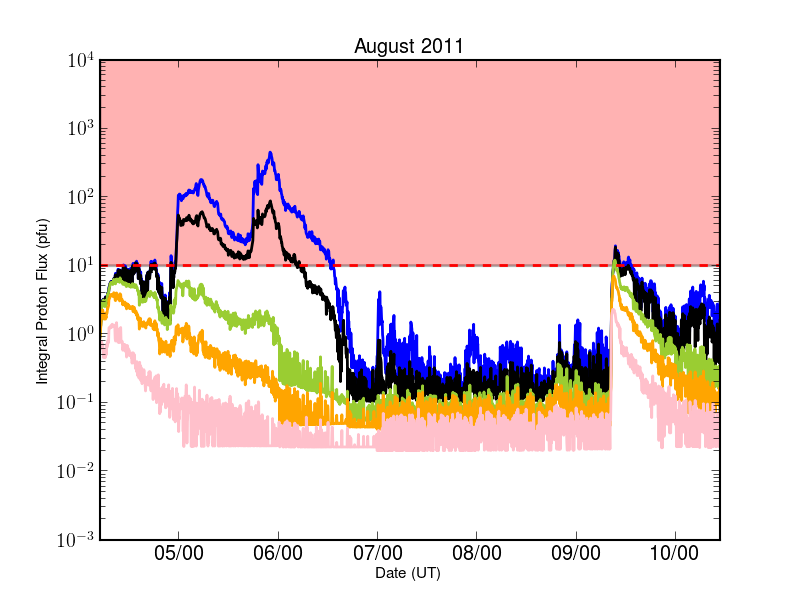}\\
\includegraphics[width=0.45\linewidth]{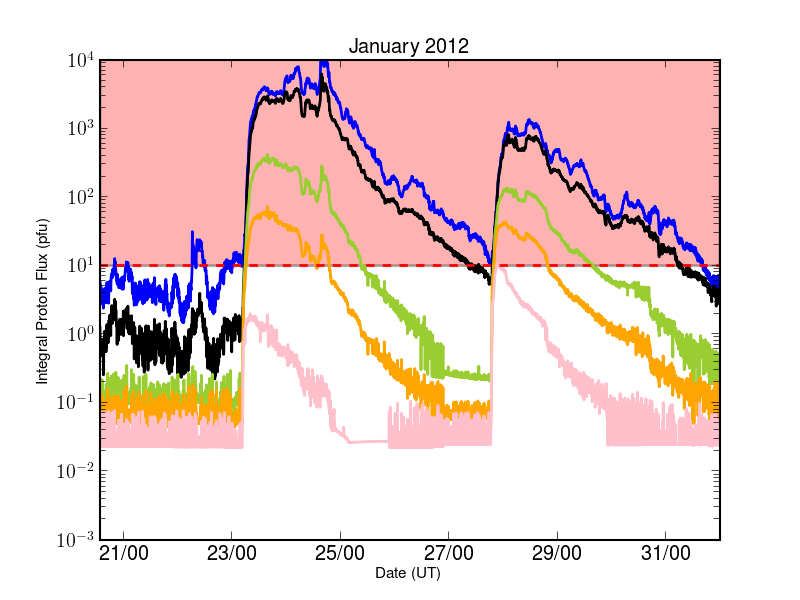}
\includegraphics[width=0.45\linewidth]{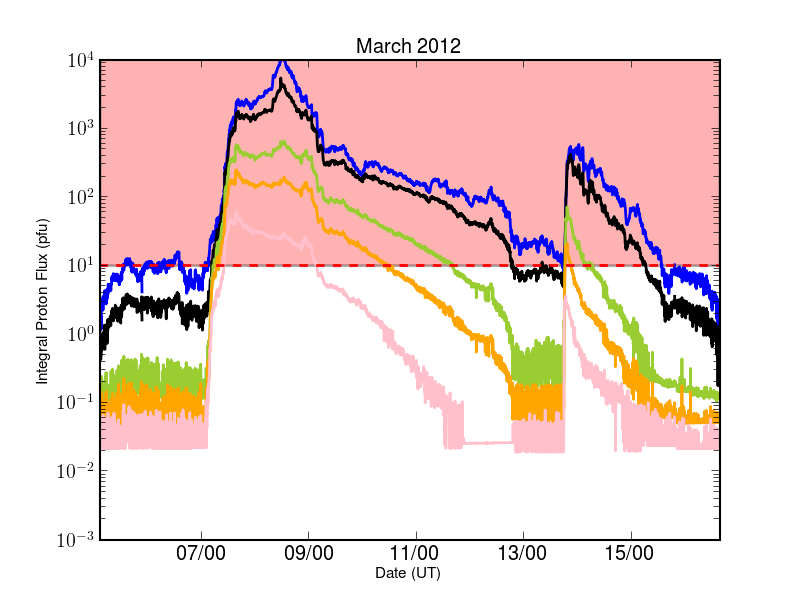}\\
\includegraphics[width=0.45\linewidth]{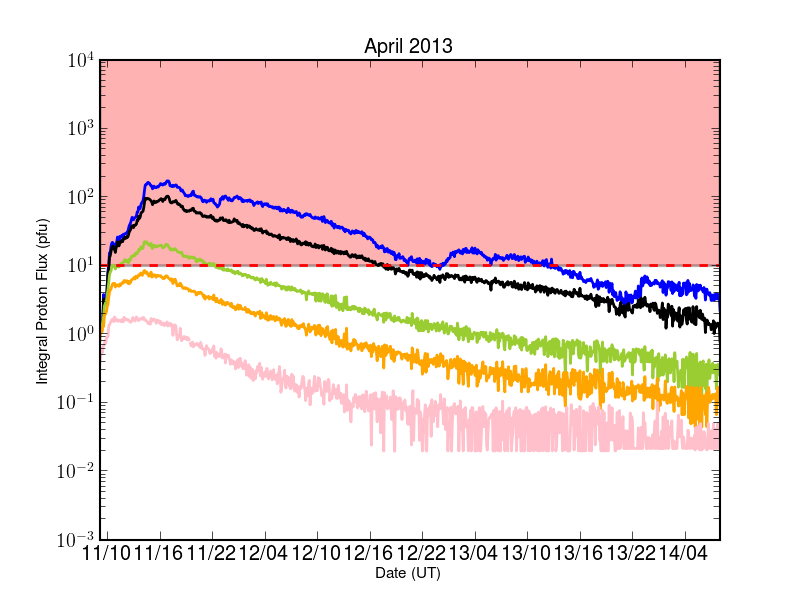}
\includegraphics[width=0.45\linewidth]{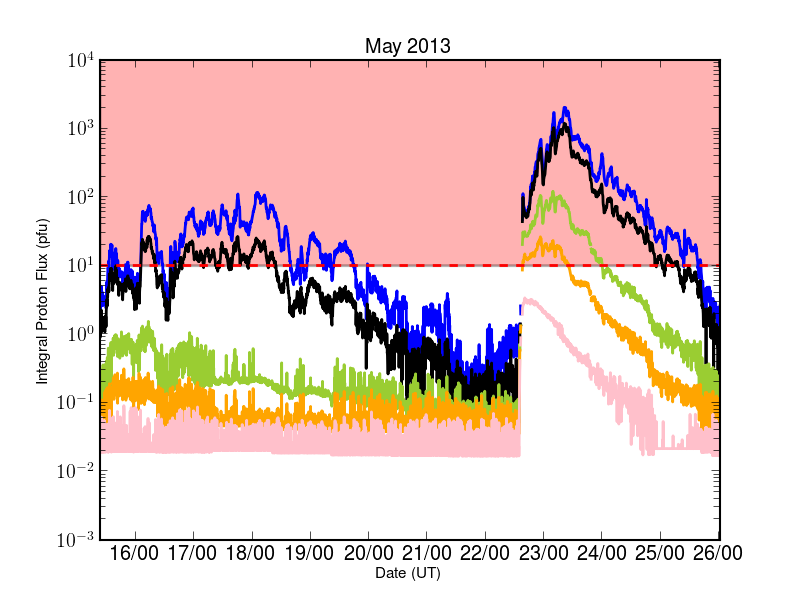}\\
\caption{Examples of solar energetic proton events, showing the NOAA warning level of $>10$ pfu in red shading.  The $> 5$\,MeV (black), $> 10$\,MeV (blue), $> 30$\,MeV (green), $> 50$\,MeV (orange), and $> 100$\,MeV (pink) proton flux are shown from the GOES-15 archived data. The day of the month is shown followed by the UT hour on the x-axis.}\label{fig-seps}
\end{center}
\end{figure}

\begin{figure}
\begin{center}
\includegraphics[width=0.45\linewidth]{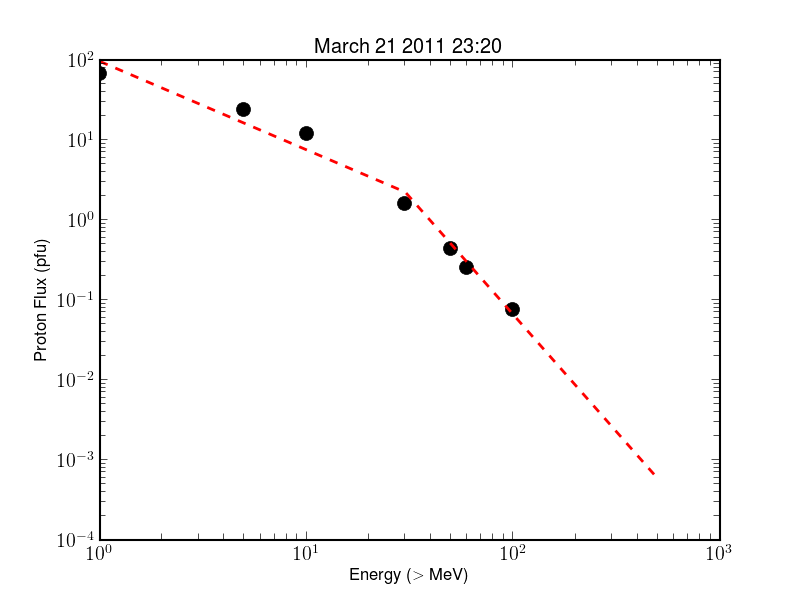}
\includegraphics[width=0.45\linewidth]{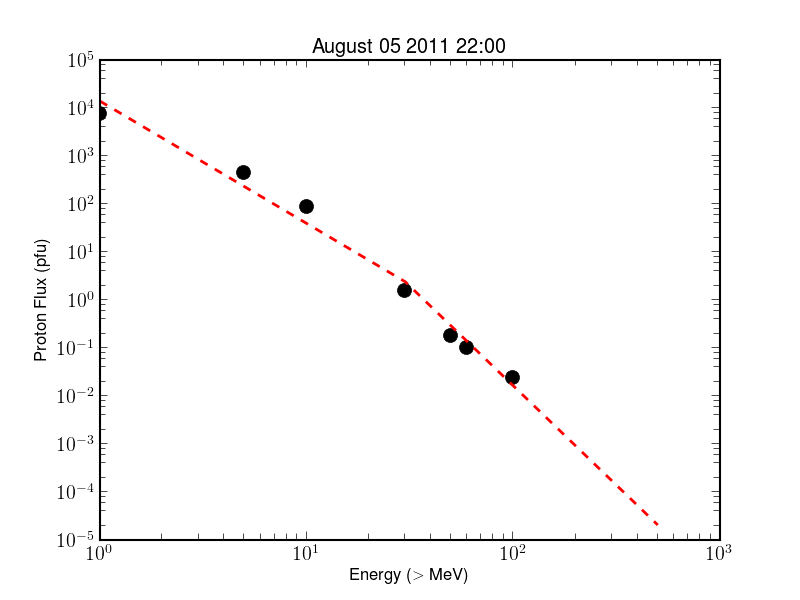}\\
\includegraphics[width=0.45\linewidth]{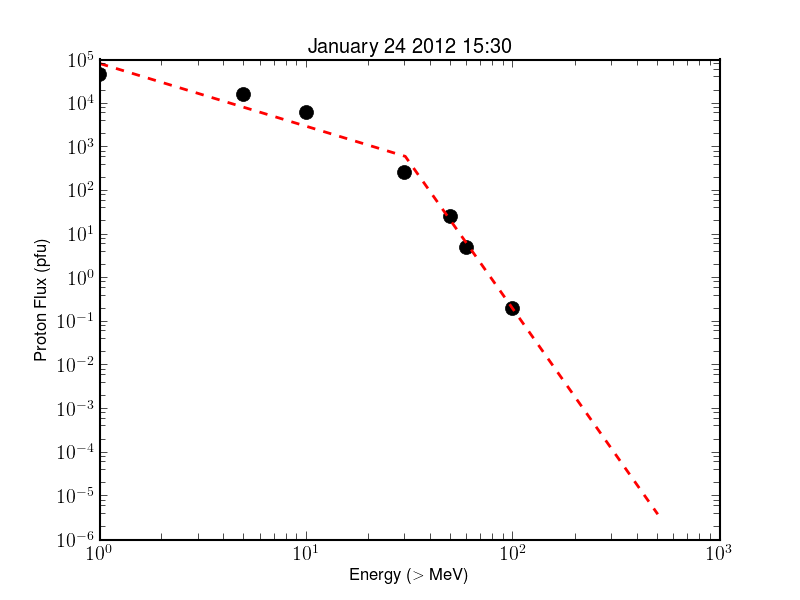}
\includegraphics[width=0.45\linewidth]{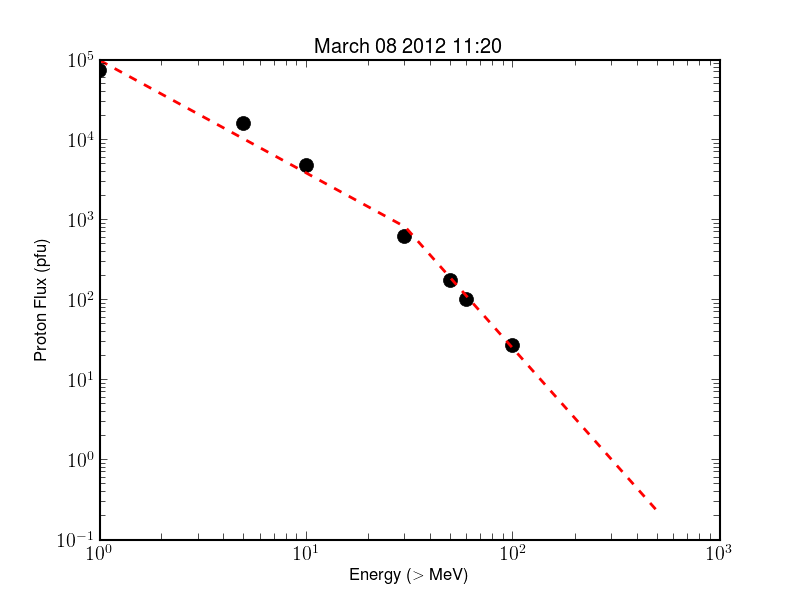}\\
\includegraphics[width=0.45\linewidth]{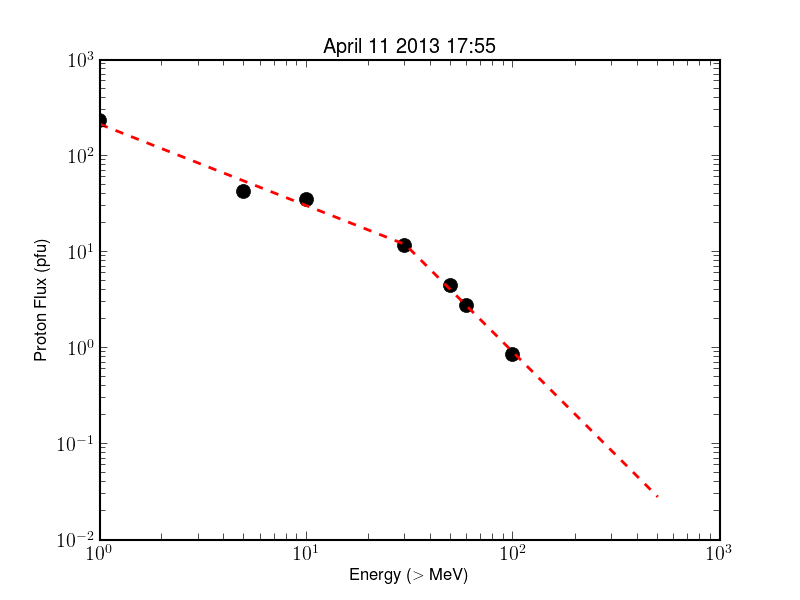}
\includegraphics[width=0.45\linewidth]{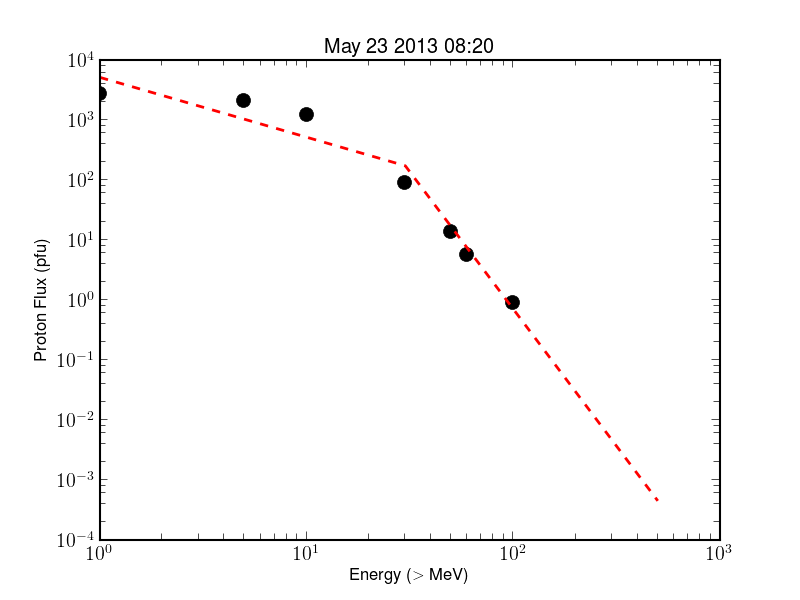}\\
\caption{Examples of the energy spectra of solar energetic proton events.  The GOES-15 energy spectra are shown with the best-fit broken power law model (dashed lines) during the peak time for the $> 10$\,MeV proton flux.  Fits are shown for the same events in Figure~\ref{fig-seps}.}\label{figure-sepfits}
\end{center}
\end{figure}

\subsection{GOES Integral Proton Data}\label{goes analysis}
Time averaged data from the GOES-13 and GOES-15 energetic proton, electron, and alpha detectors (EPEADs) were downloaded for the 2010-2013 time period.  These data include 5-min time averages of the proton flux from each of two detectors, providing East-facing and West-facing proton flux measurements.  An automatic detection algorithm was run to determine the SEP events ($> 10$\,pfu in the $> 10$\,MeV proton flux) and compared with the NOAA SWPC SEP event list.  For both the GOES-13 and GOES-15 proton flux measurements, we identify SEPs as having a peak flux at $> 10$\,MeV that is above $10$\,pfu.  The duration of the event is computed as the time from where the proton flux rises above 1\,pfu and falls below 1\,pfu on either side of the peak in the proton flux.  Since the typical background proton background level at $> 10$\,MeV is $\sim 0.1$\,pfu, we chose 1\,pfu as a value ten times above the typical background and ten times below the NOAA SWPC SEP warning level. This simple SEP event definition can result in earlier start times than those computed for the NOAA SEP list (for example, events 12 and 13 in Table~\ref{table-sepgoes13} occur 1-2 days before the NOAA start time) and also identifes closely-timed events as one event.

For each of the SEP events, we determine the onset time, peak time, event duration, peak proton flux at $> 10$\,MeV, and integrated proton flux at $> 10$\,MeV.  The median proton flux for the month when each SEP event occurs, for all measurements with the $> 10$\,MeV flux below 1\,pfu, is also determined.  These basic characteristics of the SEP events are recorded in Table~\ref{table-sepgoes13} for the GOES-13 data and Table~\ref{table-sepgoes15} for the GOES-15 data.  Examples of the proton flux in multiple integral energy bands are shown for several SEP events in Figure~\ref{fig-seps}.  These events include low flux events like that of August 2011 and the highest levels seen in recent years from January and March 2012.  These highest SEP storms, however, are still only about a quarter of the peak level from the biggest storms of the last cycle (compared to the Nov 6, 2001 and Oct 29, 2003 SEP peak levels).

Comparing the GOES-13 and GOES-15 analysis to the NOAA SWPC SEP event list, we find good agreement between the peak SEP times and levels.  The SWPC list most closely matches that of GOES-13.  However, the weak event in Oct 2011 is missing in the SWPC list (event 8). 
For the most part, the GOES-15 peak times often differ by 5-10 minutes from the GOES-13 measurements.  The SEP event durations measured are roughly consistent, such that the integrated flux is consistent except for two cases (events 6 and 25).  In both cases, the slowly declining shape of the proton flux led to larger durations in the GOES-13 calculations, but the more conservative GOES-15 values are adopted throughout the paper after visual inspection of the data.

During July 2012, several SEP events are likely associated with activity leading to the large CME of July 23 observed by STEREO on the opposite side of the Sun from Earth (e.g., \citealt{2013ApJ...770...38R}).  Four events are seen with low SEP peak levels ($\sim 10$ to $> 150$\,pfu at $> 10$\,MeV).  Due to the multiple peak structure and potential differences in the locations/pointing of the GOES-13 and GOES-15 satellites, there is a variation in the peak location and flux for SEP event 20.  A slightly higher $\sim 30$\,pfu peak is seen on the 20th by GOES-15, while the peak identification algorithm includes the 20th event as part of the SEP event whose peak is on the 18th for GOES-13.  Due to the complex SEP behavior at this time, we included event 20 by searching for sub-peaks within the longer duration event 19.  Instead, the final peak for GOES-13 is identified as the weak one on the 23rd at $\sim 13$\,pfu (corresponding to the SWPC recorded SEP event).  

To characterize the proton flux across the full integral energy range, we also determined the shape of the energy spectra at both peak flux and over the integrated SEP event.  The energy spectra were fit with both a power law and broken power law model.  The better fit was determined with the $\chi^2$ statistic.  The characterizations of the energy spectra are included in Table~\ref{table-sepeventparams}.  Example fits to the energy spectra are shown in Figure~\ref{figure-sepfits}.  The power law parameters correspond to \begin{math} {\rm F} = {\rm a} \times  {\rm E}^{\rm m}\end{math}, where F is the proton flux in pfu and E is the proton integral energy in $>$\,MeV.  The derived parameters are included in Table~\ref{table-sepeventparams}.  For the broken power law model, the characterization is similar, but with separate parameters to describe the shape at energies above and below $> 30$\,MeV (the chosen break point in the energy spectrum):  \begin{math} {\rm F}_{< 30\,MeV} = {\rm a}_1 \times  {\rm E}_{< 30\,MeV}^{{\rm m}_1}\,{\rm and}\,{\rm F}_{\ge 30\,MeV} = {\rm a}_2 \times  {\rm E}_{\ge 30\,MeV}^{{\rm m}_2}\end{math}.  The GOES-13 values are used in Table~\ref{table-sepeventparams}, with the calculated error corresponding to the difference between the GOES-13 and GOES-15 parameters for each event.

Additionally, for each of the type II solar radio bursts, we found the peak proton flux level at $> 10$\,MeV within 24 hours after the burst start time.  In all cases, an increase in proton flux is seen over the median low level SEP level.  The time and flux of the peak were measured and the energy spectrum at the peak was fit with a power law/broken power law model.  These measurements are included in Table~\ref{table-protontype2s} for all of the type II bursts.  No significant difference in the distribution of the median background proton level is seen between the high and low proton peak events.  The $\ge 10$\,pfu events have an average background of $0.17 \pm 0.22$\,pfu, while the low proton peak events ($< 10$\,pfu) have an average background of $0.16 \pm 0.18$\,pfu.

\section{SEP and non-SEP Radio Properties}\label{properties}
In \S~\ref{data analysis}, we include the radio properties measured from Wind and the proton flux properties from GOES for all 123 type II solar radio bursts recorded from 2010-2013.  During this time, twenty-seven SEP events with $> 10$\,MeV flux above 10\,pfu were recorded at Earth.  In the following section, we include a statistical analysis of the type II radio bursts and compare the properties between SEP and non-SEP associated solar radio bursts.  The results of this statistical analysis are used further in our discussion of using the results of a principal component and logistic regression analysis (\S~\ref{pca}) for forecasting SEP events (\S~\ref{forecast}).         

\subsection{Statistical Properties}
Of the 123 type II solar radio bursts occurring from 2010--2013, $24$\% are associated with an SEP event.  This is determined as the number of type II bursts with peak $> 10$\,MeV proton flux above 10\,pfu in the 24\,hr period following the start of the type II burst (30/123 from Table~\ref{table-protontype2s}; note that three radio bursts overlap such that the SEP event associated with one burst is on-going following a weaker burst within a day of the first burst). This fraction increases to 33\% (25/75) when only the bursts observed with Wind are taken into account, excluding the STEREO detected bursts associated with solar regions not facing the Earth. However, an increase in the proton flux level is seen during the timeframe of the bursts in all cases.  The SEP flux rises by at least 1.76 times the median background proton flux level and as high as $> 5000$ times during the most intense SEP event.  There are only two SEP events from our analysis (see \S\ref{goes analysis}) that are not associated with a type II radio burst from the Wind/WAVES list.  These include the 08/14/2010 and 06/16/2012 events.  Both are weaker SEP events, with peaks of 14-15\,pfu at $> 10$\,MeV, barely above the NOAA SWPC warning level.  

In this sub-section, we detail the properties of the type II and SEP events of our sample.  We investigate the radio burst data of the events in Table~\ref{table-radiointensity}, to search for distinguishing characteristics between high flux SEP events ($> 10$ MeV peak flux $> 10$\,pfu) and lower proton flux events in \S~\ref{radio-statistics}.  We examine the flare location of the events in \S~\ref{flare-statistics}.  We describe the proton integral energy spectra in \S~\ref{spectralform-statistics} and the relative timing of events in \S~\ref{timing}.  We summarize the statistics of the radio properties for SEP and non-SEP events in Table~\ref{table-summary}.

\begin{figure}
\centering
\includegraphics[width=0.9\linewidth]{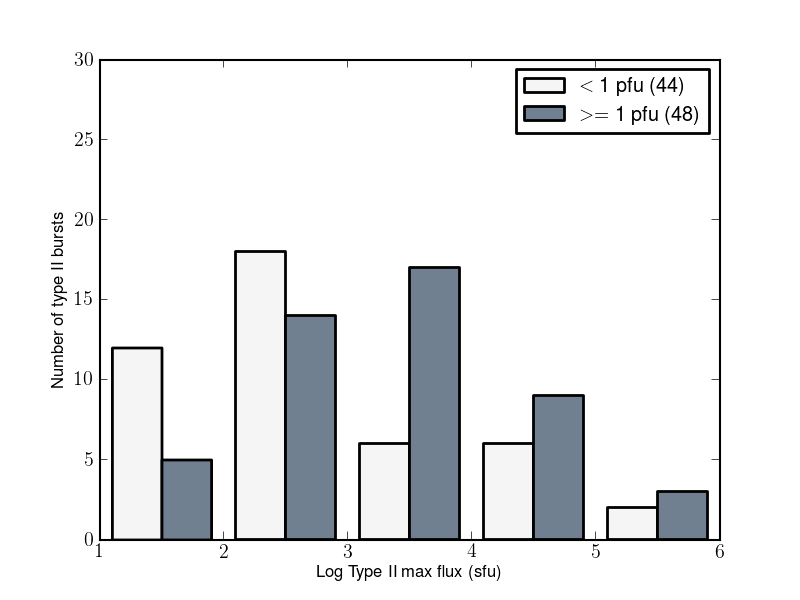}
\includegraphics[width=0.9\linewidth]{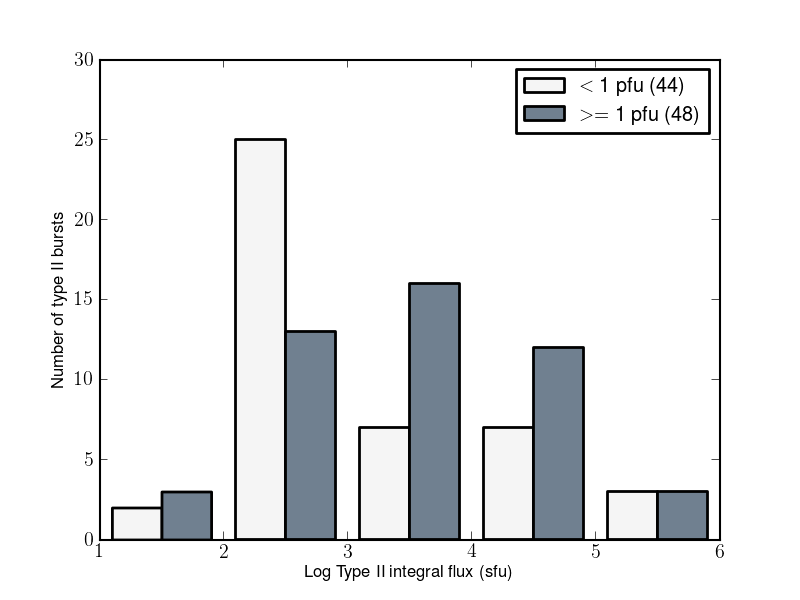}
\caption{Type II burst intensity for low and high proton flux peaks among the 92 type II bursts detected in Wind/WAVES.  For low proton flux peaks ($< 1$\,pfu at $> 10$\,MeV), both the median type II peak and integrated intensity are lower than that of the high proton flux events.  Only events visible in the Wind observations are included. { Results of a KS test give KS statistic of 0.311 and p-value of 0.018 for the comparison with the 1\,pfu threshold and a KS statistic of 0.348 and p-value of 0.005 with the 10\,pfu threshold. This indicates that the distributions of type II burst peak flux are distinct between low proton flux and high proton flux events.}  
}\label{fig-intensity}
\end{figure}

\begin{figure}
\centering
\includegraphics[width=0.95\linewidth]{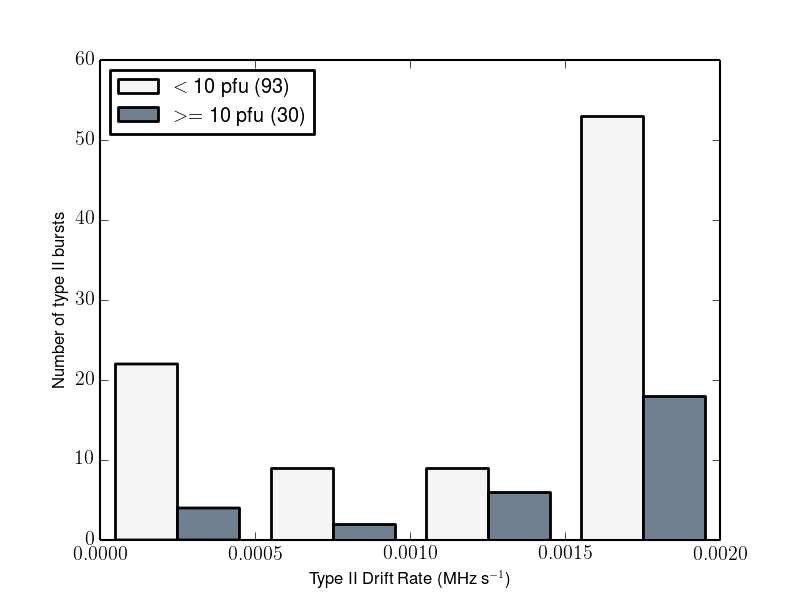}
\caption{The distribution of the absolute value of type II burst drift rates for SEP ($> 10$\,pfu) and non-SEP ($< 10$\,pfu) events.  High proton flux events are associated with high drift rates.  Conversely, none of the bursts that remain in the kilometric frequency range have high proton flux levels. { However, the KS test (KS statistic of 0.203 and p-value of 0.273) indicates no difference in the distributions for drift rates corresponding to low proton flux and high proton flux events. }
}\label{fig-drift}
\end{figure}

\subsubsection{Solar Radio Burst Characteristics}\label{radio-statistics}

{ Type II intensity, duration, and drift rate.} No direct correlation is found between the peak flux of the type II bursts and the proton peak flux or the integrated type II flux and the proton peak flux.  As found in Table~\ref{table-summary}, the median type II peak flux is two times higher and the integrated flux 0.5 times lower in the SEP events.  This is also shown in Figure~\ref{fig-intensity}, where the distributions of peak and integral type II intensity are presented for low and high proton peak flux.  { In this figure and subsequent figures (Figures 5-8), we use either the 10\,pfu threshold, corresponding to the NOAA warning level for an SEP event or the 1\,pfu threshold, an order of magnitude below the NOAA warning level and an order of magnitude above typical non-event background proton levels, to show lower proton flux events. The bin size is determined as the square-root of the sample size ($\sim 7$, but we choose from 5-7 bins depending on the range of the data).  Bins are chosen by dividing the maximum - minimum value by the bin size.}

In Figure~\ref{fig-intensity}, the flux distributions for $< 1$\,pfu and $\geq 1$\,pfu to determine whether the behavior seen at 10\,pfu also applies to less intense events. We find that more intense type II bursts are more likely to be associated with an SEP event.  The large standard deviations (a factor of 10), however, show that for any radio burst there is a great variation in the expected peak proton flux level. { Results from a Kolmogorov-Smirnov (KS) two-sample test yield a KS statistic (vertical distance between the cumulative probability distributions) of 0.311 and a two-tailed p-value of 0.018 for comparison of the integral intensities of both distributions. Comparing the peak type II burst flux distributions yields a KS statistic of 0.348 and a two-tailed p-value of 0.005. Since the p-values are low, this further shows that the low proton flux and high proton flux type II bursts have distinct distributions. }

A slight (R$^2 = 0.14$) correlation is found between the type II burst duration and the peak type II intensity (${\rm duration(hours)} \propto \log({\rm peak flux}) \times (0.035 \pm 0.011)$).  Bursts with long durations also tend to have a higher peak flux.  For the type II bursts with peaks below $10^3$\,sfu, the median duration is $\sim 1$\,hour with a standard deviation of 7.9\,hours.  The high intensity radio bursts have a much longer median duration of 8\,hours and a standard deviation of 13.7\,hours.  No direct correlation is seen, however, between the type II duration and SEP intensity or type III intensity.

The distribution of type II frequency drift rates, which are related to the shock speed, are shown in Figure~\ref{fig-drift}.  The 16\,MHz lower threshold on starting frequency necessitates that all bursts with drift rates steeper than 0.00191 MHz\,s$^{-1}$ fall in the same bin. Although this prevents us from finding a reliable median for the data set, it is qualitatively clear that SEP events are associated with higher drift rates. None of the purely-kilometric type II bursts, with absolute drift rates of $\le 0.0005$ MHz\,s$^{-1}$ were associated with proton flux $> 10$\,pfu. This agrees with the explanation put forward in \citet{2006GMS...165..207G} that the km-only type IIs are associated with CMEs that accelerate gradually and are able to drive shocks only when they have propagated far into the IP medium.  These low drift rate bursts would therefore not be able to accelerate SEPs near the Sun. In contrast, 86\% of bursts that were associated with proton flux $> 10$\,pfu had a starting frequency at or above 14 MHz (absolute drift rates of $\ge 0.00161$\,MHz\,s$^{-1}$). { However, a KS test reveals no significant difference between the drift rates of the high and low proton flux bursts, with a KS statistic of 0.203 and a two-tailed p-value of 0.273.}

\begin{figure}
\centering
\includegraphics[width=0.9\linewidth]{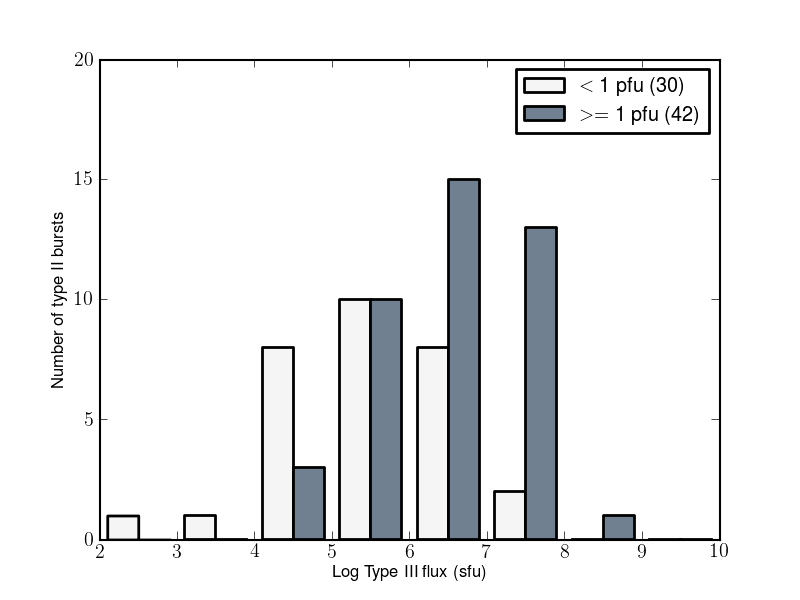}
\includegraphics[width=0.9\linewidth]{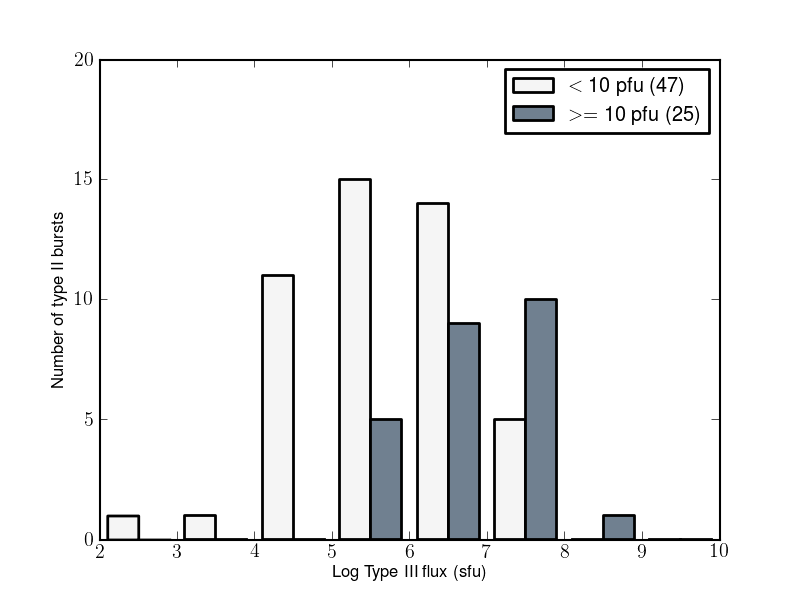}
\caption{Histogram of type III burst integral intensity for type II bursts with associated type III bursts.  For low proton flux peaks ($< 1$\,pfu at $> 10$\,MeV), the median type III integrated intensity is lower than that of the high proton flux events.  There is a much larger difference in the distributions of type III burst intensity between high and low proton flux events than was seen for the type II burst intensity in Figure~\ref{fig-intensity}. { This is indicated in the results of the KS test, with a KS statistic of 0.367 and p-value of 0.013 for the comparison with the 1\,pfu threshold and a KS statistic of 0.462 and p-value of 0.001 with the 10\,pfu threshold.} 
}\label{fig-typeIIIintensity}
\end{figure}

\begin{figure}
\centering
\includegraphics[width=0.9\linewidth]{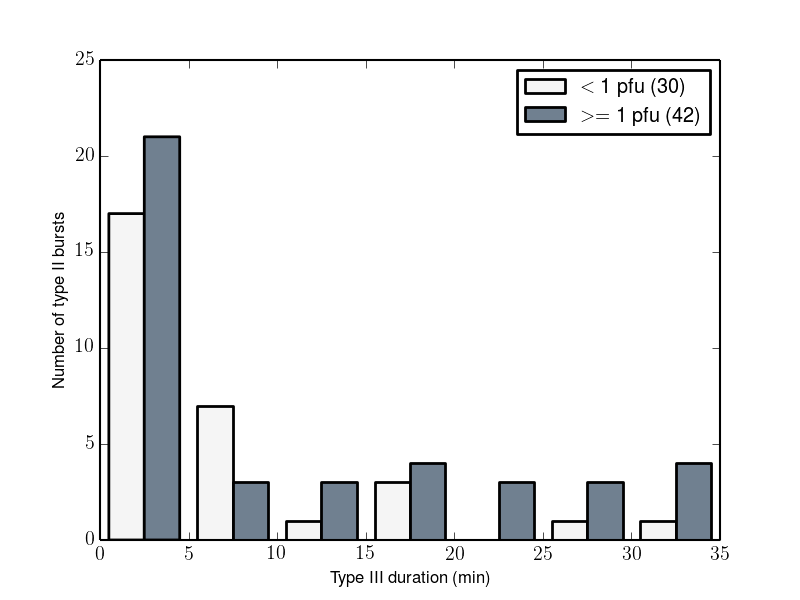}
\includegraphics[width=0.9\linewidth]{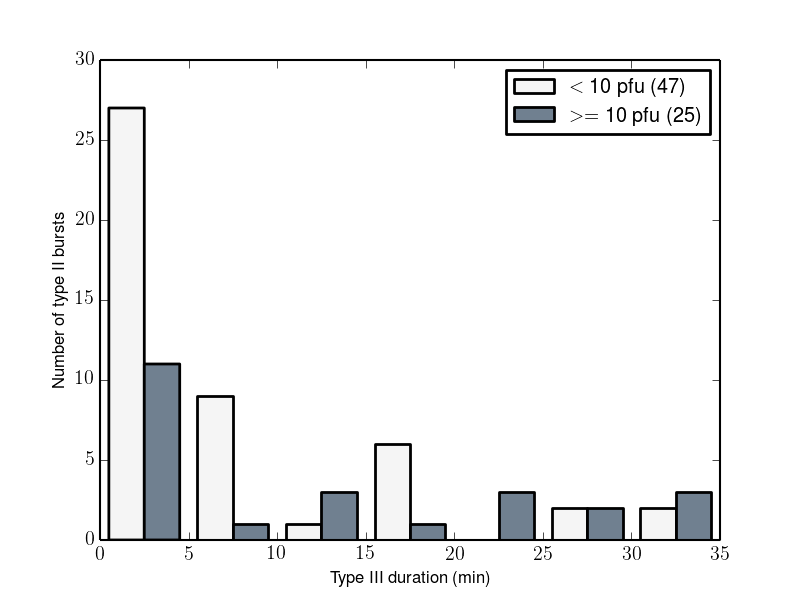}
\caption{Histogram of type III burst duration at 1\,MHz for type II bursts with associated type III burst.  The 
majority of type III burst durations are below 10 minutes for events with low proton flux peaks at $> 10$\,MeV.  The duration of higher proton flux events is longer (6\,min for $\geq 1$ pfu and 13\,min for $\geq 10$\,pfu).  As in Figure~\ref{fig-typeIIIintensity}, the number of bursts in each category is shown in the legend. { The KS test (KS statistic of 0.238 and p-value of 0.238) indicates no difference in the distributions for drift rates corresponding to low proton flux, but possible distinct populations using the 10\,pfu threshold (KS statistic of 0.286 and p-value of 0.114). 
}
}\label{fig-duration}
\end{figure}

{ Type III intensity and duration.} Type III solar radio bursts are associated with 59\% (73/123) of the type II solar radio bursts, where we take into account type III bursts occurring within 2 hrs prior to the type II burst.  Among the SEP events, this percentage increases to 92\% (25/27), which is higher than the detection rate of 57\% found for SEP events defined by their $> 30$\,MeV proton flux \citep{2009ApJ...690..598C}.  Only two SEP events are not associated with a type III burst, the 03/05/2012 and 05/15/2013 events.  For the 03/05/2012 event, the simple SEP selection criterion incorrectly identifies the start time, with the true onset occurring on 03/07/2012.  The radio bursts corresponding to \#65 in Table 1 are 4 hours before the start time listed in the NOAA SWPC list and are therefore likely associated with this SEP event. There was also a type III on 05/15/2013 at $\sim$ 2 UT but the start of the SEP event was delayed because the eruption occurred at $\sim$E64. Based on analysis of the remaining SEPs, we find that the median peak type III burst flux is $\sim 16$ times higher in the SEP events (see Table~\ref{table-summary}).  However, as for the type II bursts, the standard deviation is very large.  

Unlike the case for the distributions of type II intensity, Figure~\ref{fig-typeIIIintensity} shows that the distributions of integral type III burst intensity are more clearly separated between lower and higher intensity SEPs.  This is true for both $> 1$\,pfu and $> 10$\,pfu thresholds in proton peak flux at $> 10$\,MeV.  In terms of SEP forecasting, it is especially promising that at an integrated type III burst intensity level of $> 10^7$\,sfu,  93\% of these 17 events have a peak SEP level $> 1$\,pfu.  At the NOAA threshold of $> 10$\,pfu, this percentage is still quite high at 67\%.  Inversely, 89\% of cases where the proton flux is below 10\,pfu, the type III burst intensity is below $10^7$\,sfu.
{ The KS test shows that the distributions are distinct, with a KS statistic of 0.367 and two-tailed p-value of 0.013 using the 1\,pfu threshold and a KS statistic of 0.462 and two-tailed p-value of 0.001 using the 10\,pfu threshold.}

No direct correlation is seen between the type III burst duration and either the type III intensity, type II intensity, or SEP intensity. However, we find that the median in type III duration at 1\,MHz for higher proton flux events is at least twice that as for lower proton flux events.  Figure~\ref{fig-duration}, shows the distribution of type III duration in minutes for proton flux peaks of $< 1$\,pfu, $\geq 1$\,pfu, $< 10$\,pfu, and $\geq 10$\,pfu.  For peak proton flux at $> 10$\,MeV $>= 1$\,pfu (42 of the type II bursts), the median duration is 5.5\,min with a standard deviation of 11.2\,min, while for bursts with a measured peak flux below 1\,pfu (30 of the type II bursts), the median duration is 4.0\,min with a standard deviation of 8.2\,min.  { A KS test suggests the distributions are consistent with being drawn from the same parent population, with KS statistic of 0.238 and two-tailed p-value of 0.238.}

Using a peak proton flux threshold of 10\,pfu, it is even more apparent that longer duration bursts are associated with higher flux SEPs.  The median type III duration is 13.0\,min with a standard deviation of 11.8\,min.  The highest peak proton flux SEPs, however, do not have the longest duration type III bursts.  Of the 5 most intense SEPs, 2/5 have 18\,min or longer durations while the remaining events have type III burst durations of 10\,min or less (including the two highest peak SEPs from 01/23/2012 and 03/09/2012).  Similarly, we found that the median duration of high type III burst intensity ($\geq 10^6$\,sfu) is higher (6\,min $\pm 11$\,min) than that of weaker bursts (3\,min $\pm 8$\,min). { A KS test shows that the populations may be distinct, with a KS statistic of 0.286 and a two-tailed p-value of 0.114.}

\begin{figure}
\centering
\includegraphics[width=0.9\linewidth]{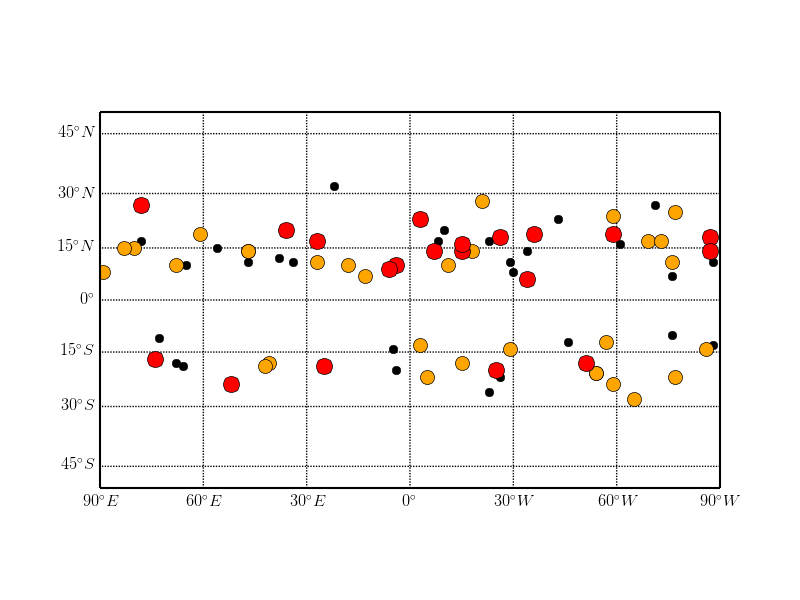}
\includegraphics[width=0.9\linewidth]{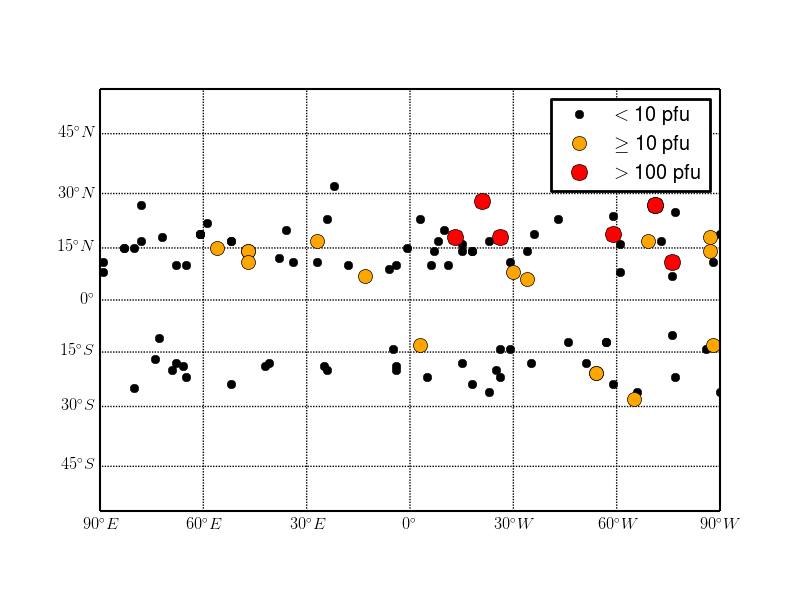}
\caption{H$\alpha$ flare location near the onset time of the DH type II radio bursts.  The top panel is color-coded by type II peak intensity, with small points for flux $< 500$\,sfu, orange for $>= 500$\,sfu, and red for $> 10^4$\,sfu.  In the bottom panel, small points are used for proton flux $< 10$\,pfu at $>10$\,MeV, orange for $> 10$\,pfu, and red for $> 100$\,pfu.  SEP events are associated with flare sites in the western hemisphere.}\label{fig-location}
\end{figure}

\begin{figure}
\centering
\includegraphics[width=0.9\linewidth]{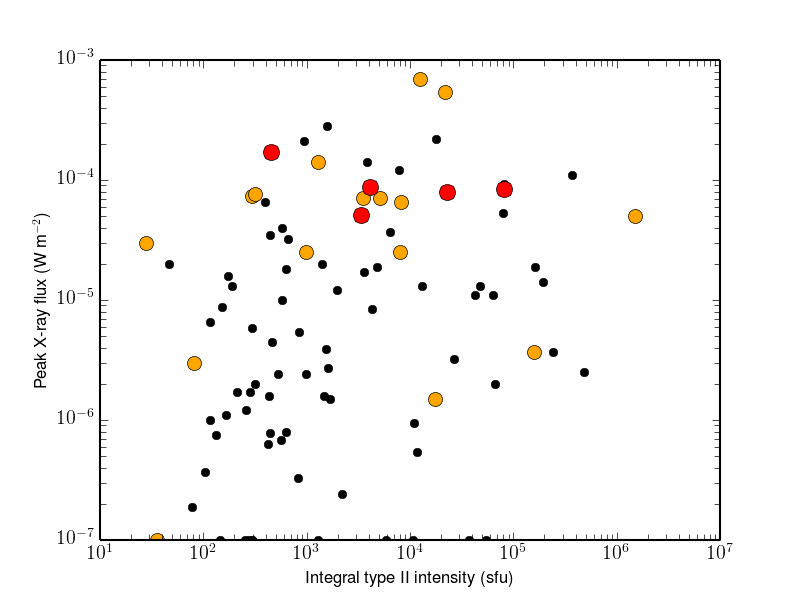}
\includegraphics[width=0.9\linewidth]{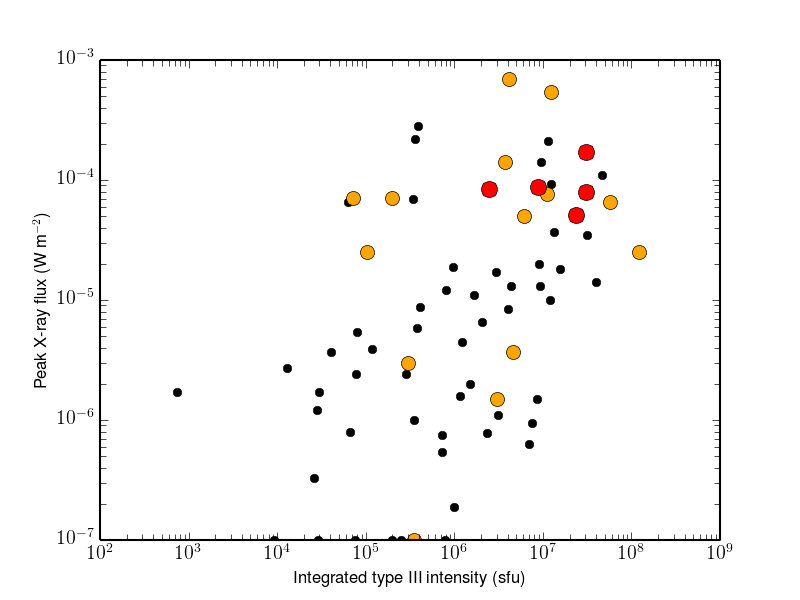}
\caption{X-ray flare class versus type II (top) and type III (bottom) integral intensity (in units of total sfu over the integration time period).  Small points are used for low proton flux $< 10$\,pfu at $>10$\,MeV, orange for $> 10$\,pfu, and red for $> 100$\,pfu, using the proton flux in the 10-hr window surrounding the type II burst start time.  In general, many of the SEP events are associated with strong X-ray flares (e.g., X-class), but there are some strong type II bursts without an associated X-ray flare.  Type III bursts show a correlation (R$^2 = 0.26$ { and Kendall tau = 0.36 with p-value of $6 \times 10^{-6}$}) with X-ray flare class.}\label{fig-xray}
\end{figure}

\subsubsection{Flare Location and X-ray Flare Class}\label{flare-statistics}
The flare location and associated X-ray flare class for each type II burst was obtained from the NOAA Space Weather Prediction Center's {\it Preliminary Report and Forecast of Solar Geophysical Data}.  Using the {\it Region Summary}, we recorded the flare information for the closest flare preceding the type II burst.  Flare location is given from H$\alpha$ optical observations in spherical, heliographic coordinates.  Flare selection included the largest flare within 12 hours preceding the type II burst.  However, if a large flare (X-class of M or greater) occurred close in time to the radio burst, this flare was selected.  Details including the flare location, class, and time difference between the start of the type II burst and the flare are included in Table~\ref{table-radiointensity}.  

Based on the flares that precede the type II bursts, the median difference between the start of the flare and the start of the type II burst is  $\sim 46$\,min with a standard deviation of 2.4\,hours.  Our criteria are less stringent than studies such as \citet{2004ApJ...605..902C}, who associate bursts with flares only when they occur no more than 15 min after the H$\alpha$ peak intensity and find 57\% of metric type II bursts associated with flares.  Other studies have found associations of metric type II bursts and flares of 79\% \citep{1975SoPh...42..445D} and 62\% \citep{1980PASAu...4...59W}.  However, our goal is different than the previous studies in that we are looking for any possible associations between flare location and radio bursts to determine the usefulness of  flare location in automatic flare prediction instead of a detailed study of the relationship between flare location and radio burst properties.

In Figure~\ref{fig-location}, the flare location is plotted for all of the type II bursts.  Among the 123 DH type II bursts, six have no flare associated with them and ten occur on the limb of the Sun (seven on the western limb).  Of the remaining 107 bursts, the flare site is associated with the active bands.  No trend is seen between the location and the magnitude of the type II burst (top panel).  However, all of the highest intensity SEP events ($> 100$\,pfu at $> 10$\,MeV) occur on locations in the west (bottom panel).  A similar result was shown for solar cycle 23 by \citet{2008AnGeo..26.3033G}, such that CMEs in the western hemisphere with associated DH type II bursts tend to produce SEP events.  This result is expected since SEPs propagate along interplanetary (IP) magnetic field lines that form Parker spirals, where the nominal magnetic connection of Earth is to the western hemisphere, while radio emission does not follow IP magnetic field lines, leading to a more even source distribution.  Based on these results, there is no obvious connection between source location and burst properties for forecasting purposes.  Though, a western flare location should correlate with a higher probability of an SEP event.

Using the associated X-ray flare class from the NOAA reports, we show the relationship between type II and type III intensity and the X-ray peak flux in Figure~\ref{fig-xray}.  No correlation is found between type II integral intensity and X-ray flare class.  However, a weak linear correlation exists between the integrated type III intensity and the X-ray peak flux as $\log{I_{\rm X-ray}} = (-11.34 \pm 0.46) + (1.02 \pm 0.07) \times \log{I_{\rm type III}}$.  Units of the X-ray peak are in W m$^{-2}$ and integrated type III intensity is in sfu (from Table~\ref{table-radiointensity}). { The correlation coefficient of $R^2 = 0.26$, indicates a correlation. Further, the Kendall's tau statistic also indicates a correlation with a tau statistic of 0.365 and a two-tailed p-value of 6 $\times 10^{-6}$.}

High X-ray flux is generally associated with SEP events, but there are cases where a strong X-ray flare is observed along with no SEP event.  Of the X-class flares associated with type II bursts, 6/15 (40\%) are not coincident with proton flux levels $> 10$\,pfu at $> 10$\,MeV.  These include bursts on 2011-02-15 (peak proton flux of 0.46 pfu), 2011-09-07 (4.7 pfu), the two type II bursts on 2011-09-22 (0.46 pfu), and the two type II bursts on 2013-05-13 (0.44 pfu).  The last two sets of bursts are associated with flares near the eastern limb.  If the near eastern limb flares (i.e., flares with a poor magnetic connection to Earth, including three flares) are excluded, then $\sim 82$\% of the X-class flares with type II bursts also show significant SEP levels reaching Earth.

\begin{figure}
\centering
\includegraphics[width=0.95\linewidth]{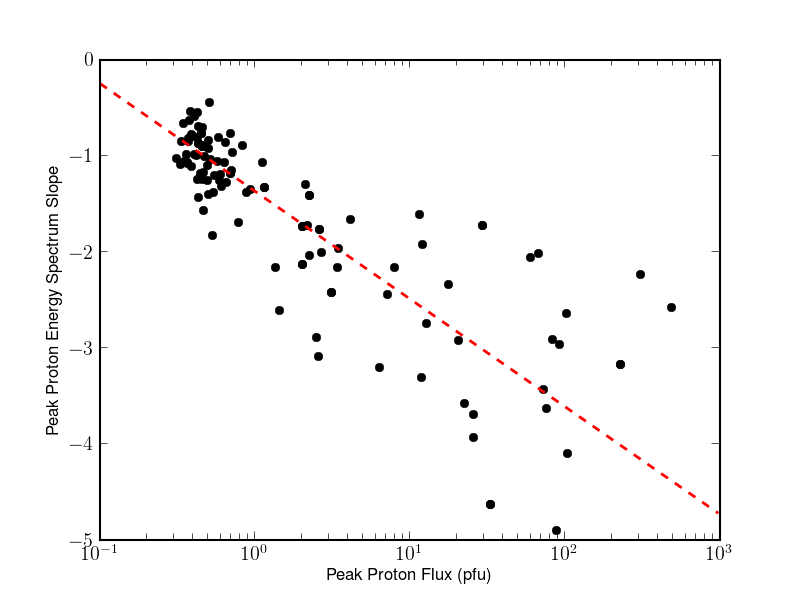}
\caption{Slope of the integral energy spectra as a function of $> 10$\,MeV proton flux.  A positive correlation is found (R$^2 = 0.69$) between the slope of the high energy spectrum (m$_{peak2}$ in Table~\ref{table-protontype2s}) and the proton flux at $> 10$\,MeV (I$_{peak}$).  This best-fit linear correlation is shown with a dashed line.  It shows the relationship of the energy spectra steepening with increasing flux.}\label{fig-peakslope}
\end{figure}

\subsubsection{Energy Spectral Form}\label{spectralform-statistics}
Examining the integral energy spectra during the peak proton flux at $> 10$\,MeV for all type II bursts, there is a definite trend of steepening spectra with higher proton flux.  In Figure~\ref{fig-peakslope}, we plot the slope of the integral energy spectra at high energies versus the peak proton flux at $> 10$\,MeV.  There is a clear correlation between the high energy slope (m$_{peak2}$ in Table~\ref{table-protontype2s}) and the proton flux (I$_{peak}$), with a linear regression fit yielding R$^2 = 0.69$.  We find that the ${\rm slope} = -1.367 \pm 0.042 + (-1.122 \pm 0.084) \times \log {\rm I_{peak}}_{> 10\,{\rm MeV}}$.   The median of the high energy slope for the lower proton flux events (I$_{peak} < 10$\,pfu) is -1.19 with a standard deviation of 0.58.  No correlation exists, however, between the lower energy slope ($< 30$\,MeV) and the peak flux.  The median slope (m$_{peak1}$) is -0.89, with a standard deviation of 0.45.

For the SEP events, there is more scatter in the range of slopes at a similar peak flux, though the correlation between proton flux and spectral form still holds (see Figure~\ref{fig-peakslope}).  This variation in integral energy form is also seen in Table~\ref{table-sepeventparams}.  The peak SEP high energy spectrum is fit with a broad range of slopes from $\sim -1$ to $-7$, with a median value of -2.83 and a standard deviation of 1.26.  The statistics on the peak flux from the 24 hrs after the type II burst are similar, with a median of -2.41 and a standard deviation of 1.11.  The slope of the energy spectrum integrated over the duration of the event is flatter with less scatter, with a median value of $-2.04$ and a standard deviation of 0.83.

\subsubsection{Timing of the Type II and Type III Bursts and the SEP Events}\label{timing}
We compared the timing of the solar radio bursts with the onset (where proton flux $> 1$\,pfu) and peak of the SEP events.  Among the 24 events associated with a type II burst (having a type II burst occur within 24 hrs of the increase in solar proton flux at $> 10$\,MeV), we find that the start of the type II burst preceded the increase in proton flux.  On average, the type II burst begins 131.3\,minutes ($\approx$ 2\,hr\,10\,min) before the $> 10$\,MeV flux increases to 10 pfu.  For the highest peak flux events ($> 100$\,pfu), the type II burst occurs from 25 min - 2 hours before the start of the SEP event. The two potential exceptions are the 01/27/2012 and 03/12/2012 events, both of which have strong bursts that occur a few days prior to the SEP event.  The type II burst onset for the weaker SEP events occurs from 20.5\,hrs before to 5.4\,hrs after the proton flux first rises above 1\,pfu. The radio bursts always occur before the peak of the SEP event. 

The peak in SEP proton flux is on average 23.7\,hours after the onset time.  However, this value ranges from a minimum of 70 minutes after the onset to 3.3\,days.  Long delays in peak intensities are due to energetic storm particle (ESP) events that occur when the CME-driven shock passes Earth.  Comparing the SEP peak time to the start of the type II burst, we find that the strongest SEP events (the 8 events with $> 100$\,pfu at $> 10$\,MeV) have, on average, a proton flux peak about 3.5\,hrs after the burst begins.  For all but two of the strong events, the SEP begins from 25 - 110 min after the burst.  The exceptions are the events from 1/27/2012 and 03/12/2012, which are events following upon higher peak events from several days prior and that have no obvious connection with a type II burst directly preceding the rise in SEP.   For less intense SEP events, the peak occurs nearly 10\,hrs after the burst (ranging to 23 hours after the burst).  Since the bursts often precede the peak by several hours, the start time of the type II burst can be useful for SEP forecasting.  However, there is a great deal of variability in the timing between the burst and SEP peak (or onset time).  More detailed investigation with a larger sample of SEP events is warranted to determine how useful the type II start time is for predicting the timing of the SEP events.     


\newpage
\section{A New Radio Index for Forecasting SEP Events}\label{pca}
The main goal of this study was to determine whether parameters measured from the dynamic radio spectra yield characteristics useful for predicting SEP events. { In the statistical analysis, we found that radio parameters such as the radio burst peak intensities had different distributions depending upon whether or not a high proton flux event was associated with the bursts. In this section, we create a new radio burst index and test whether this index is correlated with the peak proton flux following the type II radio burst. In creating a radio index, we are defining a parameter that assesses the radio burst conditions, similar to how other space weather indices indicate solar (e.g., the F10.7 index) and geomagnetic (e.g., the $Dst$ or $Kp$ indices) activity. We then will use this new parameter to compare to the proton flux with a logistic regression fit, similar to the analysis of \citet{2009SpWea...704008L}, who used the Wind/WAVES 1\,MHz radio flux as a radio index for comparison with peak X-ray flare flux to predict SEP occurrence.
}

\subsection{Principal Component Analysis to Create a New Radio Index}
We performed a principal component analysis (PCA) on the variables extracted from the radio data (e.g., intensity, duration, and slope), in order to { create a radio index that could be compared to the proton flux following a type II burst.} The analysis was done using the {\tt decompositon.PCA()} method provided in the Python library scikit-learn \citep{scikit-learn}. This analysis transforms the input vectors (here, each { radio event} is a vector in nine dimensions corresponding to the nine variables extracted from the radio data, { shown in Table~\ref{table-radiointensity}}) into a set of orthogonal components. 

The inputs of the analysis were the duration, slope, peak and integral intensity, and frequency range of the type II burst; duration, slope and integral intensity of the type III burst, if present; and the peak intensity of Langmuir waves, if present. These variables were each feature scaled to a range from 0 to 1. Variables with a range greater than 100 were log scaled first.  Feature scaling was calculated as 
\begin{math}
{\rm value}_{scaled} = ({\rm value} - min)/(max - min),
\end{math}
where the value is each measurement from an individual burst and $min$ and $max$ are the minimum and maximum of that parameter (unscaled or log-scaled) over all of the 123 bursts.  This feature scaling is a necessary step that places all of the variables on a similar scale, allowing us to determine which parameters account for the most variation in the data.

The PCA of the feature scaled variables yielded a first component (C1, { the new radio index}) that accounted for 37\% of the variance in the data. The variables that load on this component { (i.e., the relative weighting of the radio parameters that make up the radio index)} are shown in Table~\ref{table-pca}. Four variables load significantly on C1 (with weights of 0.3 or above): type III intensity and duration, type II peak intensity, and Langmuir wave peak intensity. {These variables account for the most variability in the radio measurements. The parameters with low weights, including the slope of the radio frequency profiles, type II duration, type II frequency range, and type II integral intensity, are similar between all radio events.}

{ The resulting radio index, C1, from the 9-variable PCA is:}
\begin{math}
C1 = (T_{II} \times 0.206) + (m_{II} \times 0.005) + (I_{II\,Peak}  \times 0.346) 
+ (I_{II\,Integral} \times 0.288) + (F_{II\,Range} \times 0.264) + (T_{III} \times 0.332)
 + (m_{III} \times 0.265) + (I_{III\,Integral} \times 0.614) + (I_{L\,Peak} \times 0.356). 
\end{math}  
\vspace{0.25cm}

In the above equation: $T_{II/III}$ is the duration of the type II/III burst in hours/minutes, $m_{II/III}$ is the slope of the burst, $I$ is the peak/integral intensity of type II/III bursts or Langmuir waves, and $F_{II Range}$ is the frequency range of the type II burst.

The { radio index}, $C1$, was found to be related to the occurrence of SEP events. Most of the SEP-associated type II bursts (19/23 or 83\%), and only 3 bursts without SEPs, had a $C1$ value over 0.33. This suggests that the { radio index is a} good indicator of SEP occurrence.  Therefore, using the radio parameters that make up $C1$, we can predict whether an SEP event will or will not occur. { We test this further with a logistic regression analysis in \S~\ref{sect-logreg}.}

{ The type III bursts are created from a non-thermal process during the impulsive phase of the flare, tracing electron streams propagating along open field lines (e.g., \citealt{2002JGRA..107.1315C}). The Langmuir waves are created from a secondary process, whereby the electrons from the flare site, that resulted in the type III burst, create the Langmuir waves detected at the satellite location. It is not surprising that the type III bursts are most strongly correlated with the SEP events, since the detection of type III bursts at the satellite location indicates that the satellite is magnetically well-connected to the flare site and that electrons were accelerated and escaped at a distance of  several solar radii (e.g., 1\,MHz corresponds to $\sim$ 7 $R_{\odot}$ as described in \citealt{1998SoPh..183..165L}). In the 2010-2013 sample analyzed the type II burst appears to be a necessary condition for the SEP event ($> 10$\,pfu flux for proton energies $>$ 10\,MeV), as it indicates the presence of an interplanetary CME shock. In summary, the radio parameters of most importance to the radio index are those indicating the intensity of particles accelerated towards the satellite (type III intensity and Langmuir wave peak intensity), the length of time that the accelerated particles are magnetically well-connected to the Earth (type III duration), and the intensity of the CME shock that further accelerates the particles, in the direction of the satellite (type II peak intensity). 
}

To eventually use the { radio index} as a space weather forecast tool, it is desirable to reduce the number of measured radio parameters { that make up the index to as few} as possible while still preserving their connection with { proton flux peak levels}.  Through a process of trying different combinations of the 9 variables, we found that a significant separation between SEPs and non-SEPs was preserved reducing $C1$ to the five most weighted variables from the first analysis: type III intensity and duration, type II peak intensity and integral intensity, and Langmuir wave peak intensity.  The results from this 5 component PCA are also shown in Table~\ref{table-pca}. 

Additionally, since feature scaling requires knowledge of the possible range of radio parameters future events may have, in an operational setting { this could be difficult. Instead, } it is more useful to use log-scaled variables.  We therefore determined the resultant radio index, $C1$, from a 5-component PCA performed on the log-scaled variables.  The five-variable PCA accounts for 46\% of the variance in the radio data for all of the type II bursts.  The resultant equation for the { radio index} is:

\vspace{0.25cm}

\begin{math}
C1 = T_{III} \times 0.370 + I_{II\,Peak}  \times 0.424 + I_{III\,Integral} \times 0.642 + I_{L\,Peak} \times 0.380 +  I_{II\,Integral} \times 0.356.
\end{math}
\vspace{0.25cm}

\subsection{Logistic Regression Analysis with the Radio Index and Proton Peak Flux}\label{sect-logreg}
A logistic regression analysis was { next performed to determine the probability of an SEP event occurring given the radio index. This is a similar analysis to that used in creating the SEP forecast model of \citet{2009SpWea...704008L}. By testing the effectiveness of predictions using the 9-variable and 5-variable indices, with both feature and log scaling, we will establish the effectiveness of each index for forecasting SEP events. In an operational setting, the log scaled variables are preferable, since new radio burst events could have properties with a value outside of the maximum and minimum of our current sample (the type II bursts from 2010-2013). Additionally, it is preferable to have fewer parameters for measurement in real-time (i.e., the 5-variable radio index). 

For the logistic regression analysis, the radio index (C1)} from the PCA was used as the predictor variable, with { the occurrence of an SEP event as the dependent variable. The SEP occurrence is characterized as a binary operator indicating either an SEP event (1) or no SEP event (0).} 
{ We use the scikit-learn regularized logistic regression function, {\tt sklearn.linear\_model.LogisticRegression}
\citep{yuetal-logreg}, to derive the probability of an SEP event for the sample of Wind/WAVES type II bursts using the radio index and labels of SEP or non-SEP event.} The probability calculation resulting from the logistic regression analysis is: 
\begin{equation*} P({\rm SEP | C1}) = 1/(1 + e^{-({\rm B}_0 + {\rm B}_1 \rm{C1})}).\end{equation*}  
{ Where, $P$(SEP $|$ C1) is the probability of an SEP event occurring given the radio index, C1. A probability $\ge 0.5$ indicates a prediction of an SEP, otherwise no event is predicted. The parameters B$_0$ and B$_1$ are 
determined through the regularized logistic regression fit. The best-fit values building separate models for each C1 (e.g., 9 variable feature scaled) are
listed in Table~\ref{table-logistic}. As an example of using the results of the analysis, for the type II burst on Jan 23, 2012 (\#57 in Table~\ref{table-radiointensity}) the corresponding 9-variable, feature scaled radio index is C1 =  1.80. The calculation of $P$(SEP $|$ 1.80) = 0.78, predicting an SEP event. This prediction is correct, since the radio event was associated with a significant SEP event (peak proton flux $\sim 6300$\,pfu).
}

We quantified the results of the logistic regression through the false alarm rate (FAR), percent correct score (PC), probability of detection (POD), and the Heidke Skill Score (HSS; \citealt{cite-Heidke}). 
To compute these skill scores, { we use the observed radio index and the best-fit parameters ($B_0$ and $B_1$) from the logistic regression model to make an SEP prediction for each type II burst event. We then compare the prediction to the observed occurrence/non-occurrence of an SEP, computing commonly used skill scores to assess the effectiveness of our prediction model.} The following definitions are used: the number of correct SEP event forecasts (A), the number of non-SEPs falsely forecast as an SEP event (B), the number of SEP events that occurred when no event was forecast (C), the number of correctly forecast non-SEP events (D), the number of forecasts we expect to be correct by chance (E), and the total number of forecasts (N).  Using these definitions: 
\begin{math}
{\rm FAR} =  {\rm B}/({\rm A}+ {\rm B}), 
{\rm PC} = ({\rm A} + {\rm D})/{\rm N}, 
{\rm POD} = {\rm A}/({\rm A} + {\rm C}), {\rm and}\,
{\rm HSS} = ({\rm A} + {\rm D} - {\rm E})/({\rm N} - {\rm E}).
\end{math}
The HSS is the adjusted fraction of correct forecasts, where the expected number of correct forecasts by chance was defined by \citet{2008SpWea...601001B} as:
\begin{math}
{\rm E} = [({\rm A} + {\rm B}) \times ({\rm A} + {\rm C}) + ({\rm B} + {\rm D}) \times ({\rm C} + {\rm D})]/N.
\end{math}
 These parameters are frequently used to assess the performance of SEP forecast models and a more thorough description can be found in 
\citet{2009SpWea...704008L}.  

The resulting skill parameters from our analysis are recorded in Table~\ref{table-logistic}.  The best results are found with the 9-variable feature-scaled $C1$, with a POD of 62\%, FAR of 22\%, PC of 85\%, and HSS of 0.60.  With the 5-variable log-scaled $C1$, the POD drops to 58\%, but the PC is still high (84\%) and the FAR is still low (22\%).  These statistics are extremely good, particularly in comparison to other commonly used SEP models (see Table~\ref{table-logistic}).  For instance, the SWPC Protons model obtains POD of 54\%, FAR of 42\%, and HSS of 0.55, while the \citet{2009SpWea...704008L} finds a POD of 63\%, FAR of 42\%, PC of 93\%, and HSS of 0.58.  These are models that take into account solar radio bursts, either through the occurrence of a type II burst (SWPC) or the 1 MHz flux as an indicator of bursts \citep{2009SpWea...704008L}, in addition to other properties like the X-ray flux.  In particular, we note that it is the FAR rate that separates our diagnostic from the previous models, where our 22\% is very close to the 18\% rate determined with forecaster input (see \citealt{2009SpWea...704008L}).  Therefore, using the type II and type III radio burst properties alone allow for as few false detections as found by a human forecaster.

Another important consideration for flare forecasting is the time difference between when the flare occurs and when the peak proton flux will occur.  Using the 24-hr peak proton flux, we find that the events with a high probability of being associated with an SEP event, those with $P > 0.5$, have a median peak one to two hours after the end of the type II burst.  Comparing to the beginning of the burst, we find that the 24-hour peak proton flux occurs $\sim 11$ hours after the start of the type II burst.  Further, to determine how long after the type II burst the SEP peak occurs, we subtracted the peak SEP time from the GOES 13 analysis from the end time for the type II burst for all 23 SEP events that had an associated type II burst in the Wind/WAVES data.  We found that the peak proton flux always occurred after the end of the burst, with a median difference of 12 hours and a range from 15 min later to up to 42 hours later.  
 
\section{Forecasting SEP Events with WAVES Observations}\label{forecast}
In this section, we discuss implementation of the results of the PCA and logistic regression analysis for forecasting SEP events.   The initial requirement for this forecast is availability of real-time radio observations.  The type II and type III bursts analyzed are only observable from space-accessible radio frequencies.  Ground observations will not yield the same results, since there is a much weaker connection between metric type II bursts observed from the ground and the DH type IIs studied here (\citealt{2002ApJ...572L.103G} found 71\% of metric II bursts associated with SEP events and 95\% of DH type IIs, consistent with our analysis). Currently, the Wind/WAVES data are available with a $\sim 2$\,day latency.  The average peak SEP time occurs less than a day after the onset of the DH type II burst, requiring real-time observations.  The STEREO/WAVES beacon data are available in real-time, depending on whether the satellites are in range of a ground station, but due to their locations on either side of the Sun, it is unclear how useful the radio data will be for forecasting events that will reach Earth.  This will require an analysis of the STEREO observations over the same time period, which we defer to future work.  

Assuming real-time observations are available, the first step in the forecast method is to detect a DH type II burst.  Automatic recognition of metric type II bursts, using the Hough transform to detect bursts as straight segments in dynamic spectra transformed to 1/$f$ space, has been tested with an 80\% success rate by \citet{2010ApJ...710L..58L} with the Automated Radio Burst Identification System (ARBIS).  The data, while at different frequencies, are similar to the WAVES observations and the method can easily be applied to detect type II bursts for forecasting purposes.  Factors affecting the performance of ARBIS include the strength, duration, and frequency range of the type II burst, parameters that weighted heavily in the PCA (\S~\ref{pca}).  Additional incorporation of the automated methods for detecting type III bursts \citep{2014JGRA..119..742L}, which were developed for onboard real-time detections with STEREO/WAVES, would also be applied in the forecast model.  These techniques were found to detect $\sim$ 80\% of type III bursts in testing.  For our purposes, it is the detection of a combination of a type II and a type III burst that triggers the forecast, since 92\% of the SEPs studied were accompanied by both a type II and type III burst.  When both bursts are detected, the radio parameters are next used to compute the probability of an SEP event occurring. Since the SEP peak tends to occur after the type II burst, the full flare parameters can be measured prior to determining the value of C1 from the analysis in \S~\ref{pca}.  Following the computation of C1, the logistic regression probability is calculated to determine whether an SEP enhancement is likely to occur.

Further analysis is needed, however, to refine the model beyond giving only a probability of an SEP occurring.  In many cases, the SEP will already have begun close in time to the start of the burst.  Therefore, it is important to analyze a larger sample of SEP events, in particular to determine whether an estimate of the peak flux is possible and whether further information on how long after the type II burst the SEP peak may occur.  Since Wind/WAVES observations are available starting in 1994, this allows for future analysis of an additional 102 SEP events (based on the NOAA SWPC list), increasing our sample by a factor of five.

Another issue with using the forecast method in real-time involves the only potential real-time data source, STEREO/WAVES. Over time the STEREO observations become more and more biased towards events occurring behind the Sun as the spacecraft move farther away from the Earth.  This will increase the false alarm rate by including SEPs from the backside that will not reach the Earth.  Modifications to the forecast will need to be added to include the location of the eruptive event to account for these backside events in the STEREO/WAVES observations.

\section{Conclusions}\label{conclusions}

We analyzed the space-accessible radio and proton flux data of the 123 DH type II bursts detected by Wind/WAVES occurring from 2010-2013.  We performed a statistical analysis of the catalog we developed of type II, type III, and Langmuir wave properties along with the GOES measured proton flux and spectral form for all of these events.  These radio properties were then used in principal component  and logistic regression analyses to determine whether the space-accessible radio bursts are useful for forecasting SEP events.

We found that high solar proton flux events are nearly always observed accompanying both a type II burst and a type III burst ({ as found in earlier work, e.g., \citet{2009ApJ...690..598C})}.  The type III burst intensity and duration, from type III bursts associated with a type II burst, are the most important radio properties we tested for forecasting SEPs.  There is a high correlation between the { new radio index we computed from a principal component analysis,} which is dominated by the type III intensity and duration, and the peak flux level in an SEP event.  Through the logistic regression analysis, we found that our statistical classification of SEPs and non-SEPs based solely on the radio properties has a similar percentage of false detections as those from a human forecaster at NOAA's SWPC (FAR = 22\%, see \S~\ref{pca}) and we correctly predict 85\% of type II bursts as SEP vs. non-SEP events.  

We have shown that the radio burst properties are useful in determining whether the proton peak will or will not be high and may also allow us to make estimates of the peak time.  However, the SEP onset time (the time that proton flux first rises above 1\,pfu) is not predicted in a useful way from the radio bursts, which occur often near simultaneously as the SEP onset.  Still, given the strong correlations found between the $\sim$ 0.1--16\,MHz radio observations, it is clear that they are important sources of information for forecasting SEPs.  As the Wind/WAVES data are not available in real-time, a next step in using type II/type III burst properties is to analyze the STEREO/WAVES beacon observations.

\begin{acknowledgments}
All data used in this analysis are publicly accessible from NASA (Wind/WAVES) and NOAA (GOES). The Wind/WAVES data and catalogs were obtained as follows.  The Type II and Type IV solar radio burst list is found here: \url{http://www-lep.gsfc.nasa.gov/waves/data\_products.html}.  Wind/WAVES data were obtained as IDL save files from \url{http://www-lep.gsfc.nasa.gov/waves/data\_products.html}. NOAA data and reports were obtained from the following sources.  The SWPC SEP event list is publicly available here: \url{http://swpc.noaa.gov/ftpdir/indices/SPE.txt}.  {\it The Weekly} report on Solar Geophysical Data from NOAA SWPC is available at \url{http://www.swpc.noaa.gov/weekly.html}.  GOES SEM data downloaded from here: \url{http://www.ngdc.noaa.gov/stp/satellite/goes/dataaccess.html}.  Details on the satellite and instruments are available in the {\it GOES N Series Data Book} available here: \url{http://satdat.ngdc.noaa.gov/sem/goes/goes\_docs/nop/GOES\_N\_Series\_Databook\_rev-D.pdf}.

KL gratefully acknowledges funding through NSF and the University of Colorado as part of the LASP 2013 Research Experience for Undergraduates Program in Solar and Space Physics. The authors also thank David Malaspina (LASP) for his suggestion to analyze the Langmuir waves data and Alan Ling (AER) for providing information on the X-ray flares and comments on the draft, as well as Stu Huston (AER) and Rick Quinn (AER) for useful suggestions.
\end{acknowledgments}


\newpage

\clearpage

{\footnotesize
\begin{longtable*}{c c c c c c c c c c c c c c c}
\caption{ Type II Radio Bursts from 2010-2013.\label{table-radiointensity}}\\
\hline\hline
No. & Start Date & f$_{\rm s}$ & T$_{\rm II}$ & T$_{\rm III}$ &
I$_{\rm II, I}$ & I$_{\rm III, I}$  &  I$_{\rm L, P}$  & I$_{\rm II,P}$ & $FR$ & 
m$_{\rm II}$ & m$_{\rm III}$ & FL & FP & T$_{\rm II - F}$ \\
\hline
\textbf{1} & 01/17/10 04:05 & 16 & 0.5 & 1 & 2.4 & 4.5 & -- & 2.0 & 60 & 1.9 & 11.3 & - & - & - \\ 
\textbf{2} & 03/13/10 14:00 & 0.8 & 17.0 & 0 & 3.3 & -- & 1.4 & 3.2 & 4 & 0.6 & -- & limb & B2.4 & 253.0 \\ 
\textbf{3} & 06/12/10 01:05 & 16 & 0.1 & 1 & 1.7 & 7.0 & 1.9 & 1.3 & 40 & 1.2 & 20.7 & N23W43 & M2.0 & 10.0 \\ 
\textbf{4} & 08/01/10 09:20 & ? & 8.2 & 2 & 4.4 & -- & -- & 4.4 & 13 & 0.5 & -- & N20E36 & C3.2 & 84.0 \\ 
\textbf{5} & 08/07/10 18:35 & 14 & 1.2 & 27 & 2.8 & 7.1 & -- & 2.4 & 133 & 1.5 & 12.1 & N11E34 & M1.0 & 37.0 \\ 
6 & 08/18/10 06:05 & ? & 1.7 & 7 & 2.7 & 6.1 & -- & 2.1 & 123 & -- & 14.7 & Wlimb & C4.5 & 80.0 \\ 
\textbf{7} & 08/31/10 21:00 & 16 & 0.1 & 8 & 2.5 & * & -- & 2.1 & 40 & 0.3 & 12.2 & - & - & - \\ 
\textbf{8} & 09/08/10 23:25 & 16 & 0.4 & 2 & -- & -- & -- & 4.4 & 158 & 4.6 & 13.3 & N19W90 & C3.3 & 20.0 \\ 
\textbf{9} & 01/13/11 09:15 & 16 & 0.8 & 0 & 2.0 & -- & -- & 1.7 & 125 & 0.3 & -- & Wlimb & B3.7 & 6.0 \\ 
\textbf{10} & 01/27/11 12:20 & 16 & 0.2 & 19 & 1.9 & 6.0 & -- & 1.6 & 80 & 0.2 & 18.7 & Wlimb & B1.9 & 7.0 \\ 
\textbf{11} & 01/28/11 01:15 & 16 & 0.3 & 3 & ? & -- & -- & 5.4 & 140 & 0.1 & -- & Wlimb & M1.3 & 31.0 \\ 
12 & 01/28/11 04:35 & 16 & 0.2 & 2 & ? & -- & -- & 4.5 & 145 & 1.4 & -- & Wlimb & C1.5 & 33.0 \\ 
13 & 01/31/11 16:35 & 16 & 0.6 & 0 & -- & -- & -- & 2.1 & 25 & 0.5 & 10.7 & S20E24 & B1.7 & 164.0 \\ 
\textbf{14} & 02/13/11 17:50 & 16 & 0.2 & 2 & 2.6 & 4.8 & -- & 2.5 & 80 & 1.9 & 17.1 & S20E04 & M6.6 & 22.0 \\ 
\textbf{15} & 02/15/11 02:10 & ? & 4.8 & 17 & 4.3 & 5.6 & -- & 4.0 & 156 & 0.3 & 18.3 & S18W15 & X2.2 & 26.0 \\ 
16 & 02/24/11 12:50 & ? & 2.3 & 0 & -- & -- & -- & 2.9 & 110 & 0.2 & 12.9 & N18E72 & M3.5 & 327.0 \\ 
\textbf{17} & 03/07/11 14:30 & 16 & 0.5 & 1 & 3.7 & 6.0 & -- & 3.6 & 90 & 1.4 & 18.4 & N10E18 & M1.9 & 45.0 \\ 
\textbf{18} & 03/07/11 20:00 & 16 & 12.5 & 3 & 3.8 & 7.1 & -- & 3.6 & 158 & 0.6 & 16.2 & N24W59 & M3.7 & 17.0 \\ 
\textbf{19} & 03/21/11 02:30 & 16 & 2.0 & 2 & 2.4 & 5.9 & -- & 2.0 & 151 & 0.5 & 15.8 & - & - & - \\ 
20 & 05/09/11 21:00 & 16 & 7.0 & 9 & 2.9 & 4.9 & -- & 2.5 & 151 & -- & 13.2 & NElimb & C5.4 & 18.0 \\ 
21 & 05/29/11 10:25 & 16 & 0.5 & 0 & -- & -- & 1.5 & 2.5 & 151 & -- & 13.2 & S22E65 & M1.4 & 22.0 \\ 
\textbf{22} & 05/29/11 21:10 & 16 & 16.5 & 0 & 2.2 & 5.6 & 1.5 & 2.1 & 159 & 1.4 & 10.2 & S18E68 & C8.7 & 59.0 \\ 
\textbf{23} & 06/02/11 08:00 & 15 & 0.4 & 0 & 5.4 & 4.6 & -- & 5.3 & 110 & 1.2 & 19.5 & S19E25 & C3.7 & 38.0 \\ 
\textbf{24} & 06/04/11 07:00 & 16 & 6.8 & 3 & 3.2 & 6.1 & 1.4 & 2.9 & 157 & 0.3 & 21.8 & S22W05 & C1.6 & 521.0 \\ 
\textbf{25} & 06/04/11 22:00 & 16 & 51.5 & 3 & 2.6 & 6.4 & -- & 2.3 & 159 & 1.1 & 8.8 & N17W08 & B7.8 & 328.0 \\ 
\textbf{26} & 06/07/11 06:45 & 16 & 11.2 & 29 & 3.9 & 8.1 & 2.1 & 3.8 & 157 & 0.7 & 17.1 & S21W54 & M2.5 & 29.0 \\ 
\textbf{27} & 06/07/11 18:55 & ? & 0.5 & 3 & 3.0 & 5.0 & -- & 2.7 & 41 & 1.2 & 14.8 & S21W54 & M2.5 & 759.0 \\ 
\textbf{28} & 06/13/11 04:20 & 14 & 0.5 & 0 & 2.4 & 4.5 & -- & 2.1 & 130 & 0.9 & 13.6 & N17E78 & C1.2 & 211.0 \\ 
\textbf{29} & 07/26/11 09:45 & 16 & 0.8 & 9 & 3.1 & 5.4 & -- & 2.9 & 130 & 1.3 & 13.1 & - & - & - \\ 
\textbf{30} & 08/02/11 06:15 & 16 & 1.2 & 2 & 5.3 & 7.6 & 1.6 & 5.0 & 130 & 0.4 & 32.5 & N14W15 & M1.4 & 56.0 \\ 
\textbf{31} & 08/04/11 04:15 & 13 & 36.8 & 2 & 4.9 & 7.1 & 2.0 & 4.9 & 129 & 0.7 & 14.1 & N19W36 & M9.3 & 34.0 \\ 
\textbf{32} & 08/08/11 18:10 & ? & 2.0 & 14 & 2.6 & 7.5 & 1.4 & 2.3 & 56 & 0.1 & 21.9 & N16W61 & M3.5 & 10.0 \\ 
\textbf{33} & 08/09/11 08:20 & 16 & 0.2 & 30 & 4.1 & 6.6 & -- & 4.0 & 120 & 2.2 & 22.5 & N17W69 & X6.9 & 32.0 \\ 
\textbf{34} & 09/06/11 02:00 & 14 & 21.7 & 1 & 4.9 & 7.4 & 1.6 & 4.8 & 138 & 1.0 & 29.3 & N14W07 & M5.3 & 25.0 \\ 
\textbf{35} & 09/06/11 22:30 & 16 & 17.2 & 0 & 3.0 & 7.1 & 1.0 & 3.0 & 158 & 0.7 & 16.5 & N14W18 & X2.1 & 18.0 \\ 
36 & 09/07/11 18:50 & 16 & 0.2 & 1 & ? & -- & -- & 1.9 & 11 & 0.5 & 21.4 & N14W18 & B9.1 & 26.0 \\ 
37 & 09/08/11 22:50 & 8 & 0.2 & 0 & -- & -- & -- & 1.9 & 11 & 0.5 & 21.4 & N22E59 & C2.5 & 264.0 \\ 
38 & 09/10/11 19:00 & 0.9 & 22.0 & 0 & * & * & 1.1 & 2.7 & 16 & 0.9 & 9.8 & N23E24 & C1.7 & 99.0 \\ 
\textbf{39} & 09/22/11 11:05 & 14 & 12.9 & 32 & 3.6 & 7.0 & 1.8 & 3.6 & 139 & 0.3 & 11.0 & N15E83 & X1.4 & 36.0 \\ 
40 & 09/22/11 11:15 & 16 & 1.2 & 32 & ? & ? & 1.6 & 3.7 & 85 & 2.0 & 13.5 & N15E83 & X1.4 & 46.0 \\ 
\textbf{41} & 09/24/11 12:50 & 16 & 9.9 & 15 & 3.7 & 5.3 & 1.6 & 3.3 & 157 & 0.6 & 9.7 & N14E47 & M7.1 & 17.0 \\ 
\textbf{42} & 09/24/11 13:00 & 16 & 1.2 & 15 & 3.5 & 4.9 & 1.0 & 3.5 & 80 & 0.7 & 8.5 & N14E47 & M7.1 & 27.0 \\ 
\textbf{43} & 09/24/11 19:45 & 14 & 1.5 & 12 & 1.4 & ? & 1.6 & 1.1 & 70 & 0.9 & 8.4 & N15E56 & M3.0 & 36.0 \\ 
\textbf{44} & 09/25/11 05:30 & 16 & 0.5 & 2 & 2.5 & -- & -- & 2.4 & 80 & 2.4 & -- & N11E47 & M7.4 & 59.0 \\ 
\textbf{45} & 09/29/11 19:35 & 1.5 & 0.3 & 2 & 3.2 & 4.1 & -- & 2.8 & 8 & 0.2 & 19.1 & N10W11 & C2.7 & 448.0 \\ 
46 & 10/01/11 20:45 & 16 & 0.3 & 13 & -- & * & 1.1 & 3.2 & 125 & 0.6 & 9.0 & N10W06 & M1.2 & 709.0 \\ 
\textbf{47} & 10/21/11 13:15 & 16 & 0.6 & 6 & 2.3 & 7.0 & 1.7 & 2.0 & 95 & 0.5 & 16.2 & N07W76 & M1.3 & 22.0 \\ 
\textbf{48} & 10/22/11 14:00 & 0.7 & 6.0 & 0 & 4.1 & -- & 1.2 & 3.8 & 5 & 1.1 & -- & N25W77 & M1.3 & 240.0 \\ 
\textbf{49} & 11/03/11 22:35 & ? & 2.2 & 3 & 2.5 & 5.6 & 1.3 & 2.2 & 92 & 0.6 & 10.2 & N19E61 & C5.8 & 7.0 \\ 
\textbf{50} & 11/09/11 13:30 & 16 & 3.5 & 1 & 4.6 & 6.2 & 1.1 & 4.4 & 156 & 0.5 & 16.8 & N23W03 & M1.1 & 26.0 \\ 
\textbf{51} & 11/26/11 07:15 & 10 & 40.8 & 18 & 4.2 & 6.5 & 2.6 & 4.2 & 99 & 0.6 & 24.9 & N18W87 & C1.5 & 66.0 \\ 
52 & 12/21/11 03:00 & 16 & 5.2 & 0 & -- & -- & -- & 4.2 & 99 & 0.6 & 24.9 & S19E04 & C3.1 & 3.0 \\ 
\textbf{53} & 12/24/11 11:20 & 3 & 2.5 & 1 & 2.2 & 6.5 & 1.3 & 1.9 & 23 & 0.6 & 19.3 & S19E66 & C1.1 & 16.0 \\ 
\textbf{54} & 12/25/11 18:45 & ? & 0.2 & 17 & 2.8 & ? & -- & 2.0 & 70 & 0.5 & -- & S22W26 & M4 & 34.0 \\ 
\textbf{55} & 01/02/12 15:00 & 16 & 0.8 & 8 & 2.7 & 5.5 & -- & 2.2 & 120 & 1.9 & 13.2 & Wlimb & C2.4 & 29.0 \\ 
\textbf{56} & 01/19/12 15:00 & 16 & 11.8 & 4 & 2.8 & -- & -- & 2.6 & 159 & 1.0 & -- & N32E22 & M3.2 & 76.0 \\ 
\textbf{57} & 01/23/12 04:00 & 16 & 35.0 & 20 & 3.6 & 6.9 & 2.0 & 3.6 & 159 & 1.2 & 33.6 & N28W21 & M8.7 & 22.0 \\ 
\textbf{58} & 01/27/12 18:30 & 16 & 10.2 & 30 & 2.7 & 7.5 & 1.6 & 2.6 & 158 & 1.8 & 20.2 & N27W71 & X1.7 & 53.0 \\ 
\textbf{59} & 01/27/12 18:45 & 16 & 1.6 & 30 & ? & ? & 1.6 & 2.6 & 158 & 1.8 & 20.2 & N27W71 & X1.7 & 68.0 \\ 
\textbf{60} & 02/24/12 10:30 & 0.9 & 10.5 & 0 & 3.8 & 4.0 & 1.2 & 3.6 & 8 & 1.4 & 18.9 & ? & B1 & 585.0 \\ 
\textbf{61} & 03/04/12 11:15 & 16 & 1.0 & 1 & 3.1 & * & -- & 2.9 & 80 & 0.3 & 8.9 & N19E61 & M2.0 & 46.0 \\ 
62 & 03/04/12 12:15 & 0.9 & 4.8 & 3 & ? & ? & -- & 5.4 & 158 & 0.8 & 13.7 & N19E61 & M2.0 & 106.0 \\ 
63 & 03/05/12 04:00 & 16 & 8.3 & 18 & ? & ? & 1.3 & 2.0 & 60 & 1.9 & 11.3 & N17E52 & X1.1 & 90.0 \\ 
64 & 03/05/12 04:15 & 14 & 2.8 & 18 & ? & ? & 1.3 & 3.2 & 4 & 0.6 & -- & N17E52 & X1.1 & 105.0 \\ 
\textbf{65} & 03/07/12 01:00 & 16 & 42.0 & 22 & 4.3 & 7.1 & 2.0 & 4.3 & 159 & 1.8 & 8.8 & N17E27 & X5.4 & 60.0 \\ 
66 & 03/09/12 04:10 & 14 & 1.9 & 11 & -- & -- & -- & 4.4 & 13 & 0.5 & -- & N18W13 & M6.3 & 48.0 \\ 
\textbf{67} & 03/10/12 17:55 & 14 & 18.6 & 3 & 4.9 & 6.4 & 1.3 & 4.9 & 139 & 0.3 & 14.4 & N18W26 & M8.4 & 40.0 \\ 
\textbf{68} & 03/13/12 17:35 & 16 & 6.4 & 34 & 4.4 & 7.5 & 2.2 & 4.2 & 158 & 2.0 & 18.6 & N19W59 & M7.9 & 23.0 \\ 
\textbf{69} & 03/18/12 00:20 & 16 & 1.0 & 1 & 4.7 & 6.6 & -- & 4.4 & 158 & 4.6 & 13.3 & S20W25 & M1.3 & 228.0 \\ 
\textbf{70} & 03/21/12 07:30 & 16 & 0.5 & 4 & 4.0 & 6.9 & 1.9 & 3.8 & 90 & 0.7 & 22.4 & S22W77 & B9.5 & 58.0 \\ 
71 & 03/24/12 00:40 & 16 & 10.0 & 28 & ? & * & -- & 1.7 & 125 & 0.3 & -- & S25E80 & B1 & 7.0 \\ 
\textbf{72} & 03/25/12 04:50 & 16 & 0.8 & 4 & 5.7 & * & -- & 5.4 & 140 & 0.1 & -- & S24E52 & C2.5 & 192.0 \\ 
\textbf{73} & 03/26/12 23:15 & 16 & 0.7 & 6 & 4.7 & ? &  & 4.5 & 145 & 1.4 & -- & - & - & - \\ 
\textbf{74} & 03/27/12 21:45 & ? & 0.8 & 0 & 2.5 & -- & -- & 2.1 & 25 & 0.5 & 10.7 & N20W10 & B1 & 295.0 \\ 
75 & 03/28/12 00:45 & 1 & 0.8 & 0 & -- & -- & -- & 2.1 & 25 & 0.5 & 10.7 & S14W26 & B7.9 & 27.0 \\ 
\textbf{76} & 04/07/12 19:00 & ? & 7.5 & 3 & 3.0 & 4.9 & -- & 2.6 & 2 & 0.3 & 18.6 & N17W23 & C2.4 & 121.0 \\ 
\textbf{77} & 04/09/12 12:20 & 16 & 0.7 & 1 & 3.2 & 5.1 & -- & 2.9 & 110 & 0.2 & 12.9 & N17W73 & C3.9 & 8.0 \\ 
\textbf{78} & 04/15/12 02:30 & 16 & 0.3 & 3 & 2.5 & 4.5 & -- & 2.3 & 120 & 0.8 & 15.2 & NElimb & C1.7 & 14.0 \\ 
79 & 04/17/12 02:00 & 2 & 7.0 & 0 & -- & -- &  & 3.6 & 90 & 1.4 & 18.4 & - & - & - \\ 
80 & 04/18/12 10:25 & 2.5 & 8.3 & 0 & -- & -- & 1.0 & 3.6 & 158 & 0.6 & 16.2 & S24W18 & B8.9 & 47.0 \\ 
\textbf{81} & 04/27/12 16:10 & 16 & 1.0 & 1 & 2.5 & -- & -- & 2.1 & 145 & 0.9 & 12.6 & N12E38 & C2.0 & 172.0 \\ 
82 & 05/06/12 02:55 & 1 & 3.1 & 0 & -- & -- & -- & 2.5 & 151 & -- & 13.2 & S26W90 & C1.5 & 9.0 \\ 
\textbf{83} & 05/17/12 01:40 & 16 & 4.7 & 2 & 3.5 & 7.4 & 2.7 & 3.4 & 157 & 0.9 & 23.9 & N11W76 & M5.1 & 15.0 \\ 
\textbf{84} & 05/26/12 20:50 & 16 & 2.5 & 13 & 2.6 & 6.8 & 2.7 & 2.3 & 157 & 0.4 & 16.1 & Wlimb & B6.3 & 122.0 \\ 
\textbf{85} & 06/08/12 23:10 & 16 & 0.8 & 4 & 4.6 & 5.3 & -- & 4.5 & 130 & 0.4 & 39.7 & N16W15 & B1 & 161.0 \\ 
\textbf{86} & 06/09/12 15:20 & 0.8 & 32.7 & 0 & 5.2 & -- & 1.7 & 4.9 & 5 & 0.2 & -- & S17E74 & M1.9 & 240.0 \\ 
87 & 07/02/12 08:35 & 10 & 6.1 & 1 & -- & -- & -- & 2.3 & 159 & 1.1 & 8.8 & N15E01 & M1.1 & 489.0 \\ 
\textbf{88} & 07/04/12 17:00 & 14 & 0.3 & 16 & 2.8 & 7.2 & -- & 2.5 & 60 & 0.9 & 14.2 & N14W34 & M1.8 & 27.0 \\ 
\textbf{89} & 07/05/12 22:40 & 3 & 1.2 & 0 & 2.2 & ? & -- & 1.8 & 22 & 0.8 & 12.9 & S12W46 & M1.6 & 63.0 \\ 
\textbf{90} & 07/06/12 23:10 & 16 & 4.5 & 20 & 5.6 & 7.7 & 2.3 & 5.2 & 157 & 0.4 & 26.9 & S18W51 & X1.1 & 9.0 \\ 
91 & 07/08/12 16:35 & 16 & 5.4 & 38 & 3.5 & 5.5 & -- & 3.3 & 157 & -- & 10.5 & S14W86 & M6.9 & 12.0 \\ 
\textbf{92} & 07/12/12 16:45 & 14 & 16.2 & 8 & 3.1 & 6.6 & 2.4 & 3.0 & 137 & 0.5 & 29.7 & S13W03 & X1.4 & 68.0 \\ 
\textbf{93} & 07/17/12 14:40 & ? & 14.3 & 1 & 3.6 & 6.5 & 1.9 & 3.0 & 118 & 0.2 & 30.4 & S28W65 & M1.7 & 157.0 \\ 
\textbf{94} & 07/18/12 06:15 & 16 & 0.4 & 4 & 1.9 & 5.5 & -- & 1.8 & 110 & 2.0 & 15.1 & S28W65 & C3.0 & 20.0 \\ 
\textbf{95} & 07/19/12 05:30 & ? & 0.8 & 14 & 2.5 & 7.0 & -- & 2.2 & 44 & 0.7 & 13.3 & S13W88 & M7.7 & 73.0 \\ 
\textbf{96} & 07/23/12 02:30 & 16 & 19.2 & 4 & 4.8 & 6.2 & -- & 4.8 & 159 & 0.4 & 16.4 & N27E78 & C2.0 & -531.0 \\ 
\textbf{97} & 08/12/12 16:20 & 1.5 & 0.7 & 5 & 2.8 & 4.8 & -- & 1.9 & 11 & 0.5 & 21.4 & S14E05 & B7.9 & 154.0 \\ 
98 & 08/21/12 20:30 & 10 & 1.5 & 4 & -- & -- & -- & 1.9 & 11 & 0.5 & 21.4 & ? & B3.6 & 344.0 \\ 
\textbf{99} & 08/22/12 02:00 & 2 & 5.8 & 0 & 2.9 & 4.4 & -- & 2.7 & 16 & 0.9 & 9.8 & S18E41 & B3.3 & 28.0 \\ 
\textbf{100} & 08/31/12 20:00 & 16 & 3.8 & 34 & 3.6 & 6.6 & 1.0 & 3.3 & 156 & 0.3 & 12.2 & S19E42 & C8.4 & 15.0 \\ 
\textbf{101} & 09/08/12 09:45 & 10 & 2.0 & 3 & 4.0 & 4.9 & -- & 3.7 & 85 & 2.0 & 13.5 & S14W29 & B1 & 0.0 \\ 
102 & 09/19/12 15:35 & 10 & 0.9 & 0 & -- & -- & -- & 3.7 & 85 & 2.0 & 13.5 & S20E69 & C2.6 & 31.0 \\ 
\textbf{103} & 09/20/12 15:10 & 16 & 7.8 & 10 & 2.2 & * & -- & 1.9 & 158 & 1.0 & 8.9 & S26W23 & B1 & 23.0 \\ 
\textbf{104} & 09/27/12 10:30 & 16 & 5.8 & 8 & 2.1 & 5.9 & 1.4 & 1.7 & 152 & 0.3 & 12.0 & ? & B7.5 & 217.0 \\ 
\textbf{105} & 09/27/12 23:55 & 16 & 10.3 & 1 & 5.2 & 6.7 & -- & 5.0 & 157 & 2.5 & 13.4 & N06W34 & C3.7 & 19.0 \\ 
\textbf{106} & 09/28/12 10:20 & 7 & 0.4 & 5 & 1.6 & 5.5 & -- & 1.3 & 60 & 0.5 & 17.0 & N08W30 & B1 & 54.0 \\ 
\textbf{107} & 10/14/12 00:40 & 16 & 0.6 & 4 & 3.2 & 6.9 & -- & 3.2 & 125 & 0.6 & 9.0 & N11E27 & C1.5 & 54.0 \\ 
\textbf{108} & 10/22/12 01:50 & 1 & 9.4 & 0 & 2.6 & -- & -- & 2.4 & 8 & 0.9 & -- & S11E73 & C1.6 & 13.0 \\ 
\textbf{109} & 11/23/12 23:15 & 16 & 2.2 & 0 & 2.1 & 5.5 & -- & 1.5 & 157 & 1.1 & 13.4 & N11W29 & C1.0 & -21.0 \\ 
\textbf{110} & 12/05/12 01:30 & 16 & 0.3 & 0 & 2.3 & 2.9 & -- & 1.8 & 80 & 2.9 & 23.3 & N10E65 & C1.7 & 78.0 \\ 
\textbf{111} & 02/26/13 10:20 & 8 & 1.4 & 0 & 2.8 & -- & -- & 1.9 & 70 & 0.5 & 6.8 & S10W76 & B6.8 & -262.0 \\ 
\textbf{112} & 03/05/13 03:30 & ? & 15.0 & 13 & 3.3 & 5.9 & -- & 3.1 & 70 & 0.7 & 8.8 & S12W57 & M1.2 & -257.0 \\ 
113 & 03/06/13 14:00 & 3 & 8.0 & 0 & -- & -- & -- & 4.2 & 99 & 0.6 & 24.9 & S12W57 & B8.9 & 216.0 \\ 
\textbf{114} & 03/15/13 07:00 & 14 & 14.5 & 4 & 4.8 & -- & 1.3 & 4.5 & 139 & 0.6 & 7.8 & N09E06 & M1.1 & 74.0 \\ 
\textbf{115} & 03/23/13 01:30 & 1.2 & 11.5 & 5 & 4.1 & 5.9 & -- & 3.8 & 9 & 0.3 & 16.5 & S24W59 & B5.4 & -19.0 \\ 
116 & 03/23/13 12:30 & 16 & 3.5 & 0 & ? & ? & -- & 2.0 & 70 & 0.5 & -- & S26W66 & B6.8 & 15.0 \\ 
\textbf{117} & 04/11/13 07:10 & ? & 7.8 & 4 & 3.9 & 7.8 & -- & 3.5 & 98 & 0.8 & 22.1 & N07E13 & M6.5 & 15.0 \\ 
\textbf{118} & 04/18/13 18:00 & 16 & 1.2 & 18 & 2.1 & 6.3 & 1.2 & 1.8 & 120 & 1.4 & 17.4 & N11W88 & C6.5 & 4.0 \\ 
119 & 05/01/13 02:50 & 16 & 0.3 & 7 & -- & -- & -- & 3.6 & 159 & 1.2 & 33.6 & S18W35 & C9.6 & 90.0 \\ 
120 & 05/13/13 02:20 & 16 & 0.7 & 2 & -- & -- & 1.1 & 2.6 & 158 & 1.8 & 20.2 & N11E89 & X1.7 & 27.0 \\ 
\textbf{121} & 05/13/13 16:15 & 16 & 2.9 & 27 & 3.2 & 5.6 & -- & 3.0 & 157 & 0.6 & 8.1 & N08E89 & X2.8 & 27.0 \\ 
\textbf{122} & 05/15/13 04:30 & ? & 3.0 & 0 & 3.9 & -- & -- & 3.6 & 12 & 0.9 & -- & N10E68 & X1.2 & 185.0 \\ 
\textbf{123} & 05/22/13 13:10 & 16 & 35.8 & 26 & 6.2 & 6.8 & 1.3 & 5.4 & 158 & 0.8 & 13.7 & N14W87 & M5 & 2.0 \\ 
\hline
\end{longtable*}
}
\begin{center}
\scriptsize Columns include the number of the type II radio burst (numbers in bold indicate bursts that were visible from WIND and used in the PCA), start time of the burst, starting frequency of the type II burst ($f_s$ in MHz), duration of
burst (T$_{II}$ in hours) and associated type III burst if present (T$_{III}$ in minutes), logarithm of the integrated type II 
intensity (I$_{II, I}$ in total sfu over the integration period), logarithm of the integrated type III intensity (I$_{III, I}$ in total sfu over the integration period), logarithm of the peak Langmuir wave intensity (I$_{L,P}$ in sfu), logarithm of the peak type II intensity (I$_{II,P}$ in sfu), frequency range of the type II burst ($FR$ in $10 \times$\,MHz), slope of the type II burst in 1/f space (m$_{II}$), slope of the type III burst in 1/f space (m$_{III}$), and the potential associated flare location (FL), X-ray flare peak (FP), and time difference in minutes between the start of the type II burst and flare (T$_{\rm II - F}$).
\end{center}

\clearpage

\begin{table}
\caption{ {\bf GOES-13: Solar Proton Properties for SEPs from 2010-2013.} }\label{table-sepgoes13}
\begin{center}
\footnotesize
\tabcolsep=0.11cm
\begin{tabular}{c l l c c l l c}
\hline\hline
No. & Start Date & Peak Time & Duration  & I$_{\rm peak}$ & I$_{\rm int}$  &  I$_{\rm median}$  \\
& & &  {\footnotesize days} &  {\footnotesize pfu} & {\footnotesize $\times 10^3$ pfu} & {\footnotesize pfu} \\
\hline
1 & 08/14/2010 11:35 & 08/14/2010 12:45 & 0.65 & 14.2 & 0.93 & 0.17\\\
2 & 03/07/2011 23:30 & 03/08/2011 08:00 & 3.60 & 50.4 & 17.37 & 0.15\\\
3 & 03/21/2011 05:55 & 03/22/2011 01:35 & 1.39 & 14.5 & 2.54 & 0.15\\\
4 & 06/05/2011 22:25 & 06/07/2011 18:20 & 4.45 & 72.9 & 15.57 & 0.15\\\
5 & 08/04/2011 04:45 & 08/05/2011 21:50 & 2.66 & 96.4 & 29.13 & 0.14\\\
6 & 08/09/2011 08:25 & 08/09/2011 12:10 & 1.06 & 26.9 & 2.81 & 0.14\\\
7 & 09/23/2011 03:30 & 09/26/2011 11:15 & 5.97 & 35.7 & 19.14 & 0.15\\\
8 & 10/22/2011 19:10 & 10/23/2011 15:35 & 2.13 & 13.2 & 2.20 & 0.14\\\
9 & 11/26/2011 08:45 & 11/27/2011 01:25 & 2.73 & 80.3 & 19.41 & 0.14\\\
10 & 01/23/2012 04:35 & 01/24/2012 15:30 & 4.19 & 6314.1 & 1277.51 & 0.15\\\
11 & 01/27/2012 09:10 & 01/28/2012 02:05 & 3.39 & 795.6 & 217.83 & 0.15\\\
12 & 03/05/2012 05:50 & 03/08/2012 11:15 & 7.59 & 6529.8 & 1078.88 & 0.15\\\
13 & 03/12/2012 20:05 & 03/13/2012 20:45 & 1.85 & 468.8 & 41.72 & 0.15\\\
14 & 05/17/2012 02:05 & 05/17/2012 04:30 & 1.53 & 255.4 & 25.20 & 0.14\\\
15 & 05/27/2012 00:05 & 05/27/2012 10:45 & 1.39 & 14.8 & 1.93 & 0.14\\\
16 & 06/16/2012 15:25 & 06/16/2012 22:30 & 0.56 & 14.9 & 0.98 & 0.13\\\
17 & 07/07/2012 02:25 & 07/07/2012 07:45 & 3.40 & 25.2 & 8.27 & 0.17\\\
18 & 07/12/2012 17:50 & 07/12/2012 22:25 & 2.62 & 96.1 & 16.80 & 0.18\\\
19 & 07/17/2012 16:05 & 07/18/2012 06:00 & 4.81 & 135.9 & 42.84 & 0.18\\\
20 & 07/19/2012 02:50 & 07/20/2012 04:50 & 2.42 & 81.4 & 13.15 & 0.18\\\
21 & 07/23/2012 09:05 & 07/23/2012 21:45 & 3.49 & 12.8 & 7.51 & 0.79\\\
22 & 09/01/2012 09:20 & 09/02/2012 08:50 & 3.59 & 59.9 & 15.79 & 0.15\\\
23 & 09/28/2012 01:35 & 09/28/2012 04:45 & 1.55 & 28.4 & 2.90 & 0.15\\\
24 & 03/16/2013 03:35 & 03/17/2013 07:00 & 1.84 & 16.0 & 3.57 & 0.13\\\
25 & 04/11/2013 09:15 & 04/11/2013 16:45 & 2.93 & 113.6 & 18.60 & 0.14\\\
26 & 05/15/2013 10:00 & 05/17/2013 17:20 & 5.48 & 41.7 & 22.66 & 0.13\\\
27 & 05/22/2013 14:20 & ... & ... &... & ... & ... \\
\hline
\end{tabular}
\end{center}

\begin{center}
\scriptsize Derived measurements from the GOES-13 data for all SEP events from 2010 - May 2013.  Columns include the number of the SEP event, start time (where proton flux rises above 1\,pfu), peak time, duration of
the event in days, the peak intensity (I$_{\rm peak}$ in pfu), the integrated intensity (I$_{\rm int}$ in pfu), and the median over the month when an SEP event is not occurring (I$_{\rm median}$ in pfu).  Data were missing from the GOES-13 proton flux files during the period covering the May 23, 2013 SEP event.  For all but event 20, which is taken from the West-facing measurement, the values are from the East-facing detector. 
\end{center}
\end{table}

\clearpage
\begin{table}
\caption{{\bf GOES-15: Solar Proton Properties for SEPs from 2010-2013.}}\label{table-sepgoes15}
\begin{center}
\footnotesize
\tabcolsep=0.11cm
\begin{tabular}{c l l l c l l c}
\hline\hline
No. & Start Date & Peak Time & Duration & I$_{\rm peak}$ & I$_{\rm int}$  &  I$_{\rm median}$  \\
& & &  {\footnotesize days} &  {\footnotesize pfu} & {\footnotesize $\times 10^3$ pfu} & {\footnotesize pfu} \\
\hline
1 & 08/14/2010 11:25 & 08/14/2010 12:55 & 0.70 & 15.5 & 0.98 & 0.17\\\
2 & 03/07/2011 23:15 & 03/08/2011 08:05 & 3.81 & 50.8 & 17.59 & 0.15\\\
3 & 03/21/2011 18:00 & 03/21/2011 23:20 & 0.86 & 11.9 & 1.31 & 0.15\\\
4 & 06/07/2011 07:10 & 06/07/2011 18:15 & 2.15 & 67.2 & 10.23 & 0.16\\\
5 & 08/04/2011 04:55 & 08/05/2011 22:00 & 2.38 & 88.1 & 13.67 & 0.14\\\
6 & 08/09/2011 08:20 & 08/09/2011 09:20 & 0.47 & 18.7 & 0.88 & 0.14\\\
7 & 09/23/2011 04:10 & 09/26/2011 12:00 & 5.40 & 37.4 & 18.80 & 0.15\\\
8 & 10/22/2011 19:35 & 10/23/2011 15:30 & 2.08 & 12.3 & 1.97 & 0.14\\\
9 & 11/26/2011 09:30 & 11/27/2011 01:20 & 2.69 & 67.8 & 17.41 & 0.15\\\
10 & 01/23/2012 04:25 & 01/24/2012 15:30 & 4.19 & 6263.6 & 1271.54 & 0.15\\\
11 & 01/27/2012 09:00 & 01/28/2012 02:20 & 3.41 & 807.5 & 209.55 & 0.15\\\
12 & 03/05/2012 05:55 & 03/08/2012 11:20 & 7.53 & 5456.1 & 1071.59 & 0.14\\\
13 & 03/12/2012 18:45 & 03/13/2012 20:45 & 2.50 & 421.7 & 45.72 & 0.14\\\
14 & 05/17/2012 02:10 & 05/17/2012 03:55 & 1.16 & 181.6 & 10.54 & 0.14\\\
15 & 05/27/2012 00:20 & 05/27/2012 10:25 & 0.99 & 12.2 & 1.24 & 0.19\\\
16 & 06/16/2012 16:35 & 06/16/2012 20:20 & 0.57 & 14.8 & 0.69 & 0.13\\\
17 & 07/07/2012 01:50 & 07/07/2012 07:30 & 3.31 & 20.9 & 7.17 & 0.25\\\
18 & 07/12/2012 19:25 & 07/12/2012 22:25 & 2.53 & 75.4 & 11.78 & 0.25\\\
19 & 07/17/2012 15:55 & 07/18/2012 06:00 & 4.32 & 104.5 & 34.32 & 0.25\\\
20 & 07/19/2012 06:25 & 07/20/2012 02:45 & 1.99 & 29.1 & 2.75 & 0.17\\\
21 & 07/23/2012 10:00 & 07/24/2012 03:35 & 2.16 & 10.3 & 3.07 & 0.29\\\
22 & 09/02/2012 00:50 & 09/02/2012 12:50 & 2.47 & 41.5 & 9.51 & 0.14\\\
23 & 09/28/2012 01:35 & 09/28/2012 04:50 & 1.59 & 23.0 & 2.74 & 0.14\\\
24 & 03/16/2013 03:50 & 03/17/2013 07:05 & 1.81 & 13.8 & 3.27 & 0.13\\\
25 & 04/11/2013 09:20 & 04/11/2013 17:55 & 1.37 & 35.0 & 5.68 & 0.13\\\
26 & 05/15/2013 09:15 & 05/16/2013 05:15 & 1.42 & 26.8 & 3.19 & 0.13\\\
27 & 05/22/2013 14:20 & 05/23/2013 08:20 & 2.33 & 1196.9 & 167.39 & 0.13\\
\hline
\end{tabular}
\end{center}

\begin{center}
\scriptsize As for Table~\ref{table-sepgoes13}, the columns have the same format, but correspond to measurements from the GOES-15 data.  Columns include the number of the SEP event, start time (where proton flux rises above 1\,pfu), peak time, duration of
the event, logarithm of the peak intensity (I$_{\rm peak}$ in pfu), logarithm of the integrated intensity (I$_{\rm int}$ in pfu), and logarithm of the median over the month when an SEP event is not occurring (I$_{\rm median}$ in pfu).  The majority of the SEPs values are from the East-facing detector.  However, events 15, 17, 18, and 19 are from the West-facing detector, which shows better agreement with the GOES-13 measurements (e.g., flare 15 is not seen in the East-facing detector and the timing of the July 2012 events is somewhat different.).  Event 22 is listed from the East-facing detector, but in the West-facing detector it begins at 09/01/2012 08:30, with the same peak time, but reaches a lower peak flux (33.6\,pfu).
\end{center}
\end{table}

\clearpage

\begin{table}
\caption{Spectral Energy Fits for SEP Events from GOES-13.}\label{table-sepeventparams}
\begin{center}
\footnotesize
\tabcolsep=0.11cm
\begin{tabular}{c c c c c c c c c c}
\hline\hline
{\bf No.} & {\bf Peak Date} & {\bf a$_{\rm Peak 1}$} & {\bf m$_{\rm Peak 1}$} & {\bf a$_{\rm Peak 2}$} & {\bf m$_{\rm Peak 2}$}  &  \bf a$_{\rm Int 1}$ & \bf m$_{\rm Int 1}$ & \bf a$_{\rm Int 2}$ & \bf m$_{\rm Int 2}$\\
\hline

1 & 08/14/2010 & 1.7 $\pm$ 1.6 & -0.99 $\pm$ 0.05 & 5.7 $\pm$ 5.1 & -3.69 $\pm$ 0.45 & 1.8 $\pm$ 1.8& -1.35 $\pm$ 0.01& 3.2 $\pm$ 3.1& -2.29$\pm$ 0.06\\ 
2 & 03/08/2011 & 2.5 $\pm$ 2.5 & -1.08 $\pm$ 0.15 & 8.1 $\pm$ 7.1 & -4.84 $\pm$ 0.56 & 2.7 $\pm$ 2.7& -1.74 $\pm$ 0.02& 4.5 $\pm$ 4.4& -2.98$\pm$ 0.10\\ 
3 & 03/22/2011 & 2.2 $\pm$ 2.0 & -3.00 $\pm$ 0.07 & 4.8 $\pm$ 4.7 & -1.22 $\pm$ 0.12 & 0.9 $\pm$ 2.7& -0.99 $\pm$ 0.09& 0.6 $\pm$ 4.5& -1.21$\pm$ 0.03\\ 
4 & 06/07/2011 & 2.6 $\pm$ 2.4 & -0.80 $\pm$ 0.10 & 4.4 $\pm$ 4.0 & -2.00 $\pm$ 0.14 & 2.0 $\pm$ 2.0& -0.96 $\pm$ 0.08& 3.2 $\pm$ 3.2& -1.72$\pm$ 0.12\\ 
5 & 08/05/2011 & 4.0 $\pm$ 3.6 & -2.36 $\pm$ 0.18 & 6.8 $\pm$ 6.5 & -4.27 $\pm$ 0.12 & 2.9 $\pm$ 2.8& -1.44 $\pm$ 0.16& 4.9 $\pm$ 3.9& -2.77$\pm$ 0.43\\ 
6 & 08/09/2011 & 2.0 $\pm$ 1.6 & -2.29 $\pm$ 0.77 & 4.4 $\pm$ 3.3 & -0.72 $\pm$ 0.40 & 0.9 $\pm$ 2.9& -1.05 $\pm$ 1.05& 0.8 $\pm$ 4.9& -1.13$\pm$ 1.13\\ 
7 & 09/26/2011 & 2.4 $\pm$ 0.5 & -1.97 $\pm$ 0.15 & ... & ... & 2.7 $\pm$ 2.6& -1.85 $\pm$ 0.03& 3.8 $\pm$ nan& -2.61$\pm$ 0.17\\ 
8 & 10/23/2011 & 1.4 $\pm$ 0.2 & -1.52 $\pm$ 0.24 & ... & ... & 3.0 $\pm$ 3.0& -2.51 $\pm$ 0.00& 1.6 $\pm$ 1.6& -1.53$\pm$ 0.02\\ 
9 & 11/27/2011 & 3.1 $\pm$ 2.8 & -1.61 $\pm$ 0.11 & 7.6 $\pm$ 7.5 & -4.70 $\pm$ 0.04 & 3.0 $\pm$ 3.0& -2.00 $\pm$ 0.01& 4.7 $\pm$ 4.7& -3.15$\pm$ 0.01\\ 
10 & 01/24/2012 & 4.9 $\pm$ 4.8 & -1.39 $\pm$ 0.05 & 12.6 $\pm$ 12.5 & -6.63 $\pm$ 0.08 & 3.9 $\pm$ 3.8& -1.08 $\pm$ 0.00& 10.1 $\pm$ 9.9& -5.29$\pm$ 0.08\\ 
11 & 01/28/2012 & 3.5 $\pm$ 3.4 & -2.96 $\pm$ 0.19 & 6.7 $\pm$ 5.9 & -0.77 $\pm$ 0.05 & 2.2 $\pm$ 3.9& -2.94 $\pm$ 0.03& 5.2 $\pm$ 10.1& -0.88$\pm$ 0.01\\ 
12 & 03/08/2012 & 5.0 $\pm$ 5.0 & -1.35 $\pm$ 0.02 & 7.4 $\pm$ 7.3 & -2.99 $\pm$ 0.04 & 3.6 $\pm$ 3.6& -0.96 $\pm$ 0.01& 5.7 $\pm$ 5.6& -2.39$\pm$ 0.06\\ 
13 & 03/13/2012 & 3.2 $\pm$ 3.1 & -3.54 $\pm$ 0.20 & 7.1 $\pm$ 6.8 & -0.87 $\pm$ 0.07 & 1.9 $\pm$ 3.6& -2.57 $\pm$ 0.04& 3.8 $\pm$ 5.7& -1.28$\pm$ 0.03\\ 
14 & 05/17/2012 & 2.7 $\pm$ 2.4 & -0.41 $\pm$ 0.19 & 4.8 $\pm$ 4.6 & -1.85 $\pm$ 0.10 & 2.7 $\pm$ 2.1& -0.92 $\pm$ 0.27& 3.8 $\pm$ 3.2& -1.71$\pm$ 0.32\\ 
15 & 05/27/2012 & 1.7 $\pm$ 0.0 & -1.71 $\pm$ 0.00 & ... & ... & 1.2 $\pm$ 0.0& -1.04 $\pm$ 0.00& 0.6 $\pm$ 0.0& -1.45$\pm$ 0.00\\ 
16 & 06/16/2012 & 1.7 & -1.69 & ... & ... & 3.1 $\pm$ 3.0& -2.48 $\pm$ 0.10& 2.3 $\pm$ 1.9& -1.93$\pm$ 0.20\\ 
17 & 07/18/2012 & 3.1 $\pm$ 0.0 & -1.34 $\pm$ 0.00 & 8.0 $\pm$ 0.0 & -4.69 $\pm$ 0.00 & 2.6 $\pm$ 0.0& -1.34 $\pm$ 0.00& 5.8 $\pm$ 0.0& -3.47$\pm$ 0.00\\ 
18 & 07/12/2012 & 2.9 $\pm$ 0.0 & -4.19 $\pm$ 0.00 & 6.9 $\pm$ 0.0 & -1.53 $\pm$ 0.00 & 2.2 $\pm$ 0.0& -2.10 $\pm$ 0.00& 3.1 $\pm$ 0.0& -1.49$\pm$ 0.00\\ 
19 & 07/19/2012 & 2.7 $\pm$ 2.3 & -0.92 $\pm$ 0.08 & 6.5 $\pm$ nan & -3.53 $\pm$ 1.26 & 2.6 $\pm$ 2.0& -1.32 $\pm$ 0.36& 5.6 $\pm$ 4.1& -3.33$\pm$ 0.93\\ 
20 & 07/07/2012 & 2.1 $\pm$ 0.0 & -2.95 $\pm$ 0.00 & 5.2 $\pm$ 0.0 & -0.83 $\pm$ 0.00 & 1.5 $\pm$ 0.0& -2.07 $\pm$ 0.00& 3.1 $\pm$ 0.0& -0.97$\pm$ 0.00\\ 
21 & 07/23/2012 & 2.0 $\pm$ 0.0 & -0.84 $\pm$ 0.00 & 3.3 $\pm$ 0.0 & -1.73 $\pm$ 0.00 & 1.6 $\pm$ 0.0& -0.68 $\pm$ 0.00& 3.8 $\pm$ 0.0& -2.20$\pm$ 0.00\\ 
22 & 09/02/2012 & 2.0 $\pm$ 0.1 & -1.87 $\pm$ 0.03 & ... & ... & 1.6 $\pm$ 0.1 & -1.59 $\pm$ 0.01& ... & ...\\ 

23 & 09/28/2012 & 2.0 $\pm$ 2.0 & -3.51 $\pm$ 0.32 & 5.9 $\pm$ 5.2 & -0.88 $\pm$ 0.13 & 1.2 $\pm$ 8.2& -1.12 $\pm$ 0.07& 0.8 $\pm$ 1.7& -1.44$\pm$ 0.15\\ 
24 & 03/17/2013 & 4.0 $\pm$ 3.9 & -3.07 $\pm$ 0.00 & 3.0 $\pm$ 2.9 & -2.41 $\pm$ 0.03 & 3.1 $\pm$ 3.0& -2.40 $\pm$ 0.01& 2.6 $\pm$ 2.6& -2.10$\pm$ 0.01\\ 
25 & 04/11/2013 & 2.7 $\pm$ 2.3 & -0.79 $\pm$ 0.06 & 5.2 $\pm$ 4.3 & -2.52 $\pm$ 0.37 & 2.3 $\pm$ 2.1& -1.05 $\pm$ 0.14& 4.2 $\pm$ 3.3& -2.33$\pm$ 0.58\\ 
26 & 05/17/2013 & 2.8 $\pm$ 0.0 & -2.08 $\pm$ 0.00 & ... & ... & 2.8 $\pm$ 0.0& -2.00 $\pm$ 0.00& 4.2 $\pm$ 0.0& -2.96$\pm$ 0.00\\ 
   27 & 05/23/2013 & 3.7 & -1.00 & 9.0 & -4.57 & 3.1 & -0.95 & 6.9 & -3.51\\ 
\hline \\
\end{tabular}
\end{center}

\begin{center}
\scriptsize Parameter fits from a single power law or broken power law fit to the peak energy spectrum (Peak) and the integral energy spectrum over the duration of the event (Int).  The constant is denoted with {\bf a} and the slope is denoted with {\bf m}.  Where a power law is the best fit, the second set of parameters are not indicated.  Error bars are derived from comparing the GOES-13 and GOES-15 results.  Large differences exist between the GOES-13 and GOES-15 results for the low energy spectrum fit to event 16.
\end{center}
\end{table}

\clearpage

{\footnotesize
\begin{longtable*}{p{3mm} p{15mm} p{8mm} p{8mm} p{8mm} p{8mm} p{8mm} p{10mm} | p{3mm} p{15mm} p{8mm} p{8mm} p{8mm} p{8mm} p{8mm} p{8mm}}
\caption{ Proton Flux Properties for Type II Radio Bursts.}\label{table-protontype2s}
\\
\hline
No. & Peak Date & I$_{\rm median}$ & I$_{\rm peak}$ &{\bf a$_{\rm Peak 1}$} & {\bf m$_{\rm Peak 1}$}  &  {\bf a$_{\rm Peak 2}$} & {\bf m$_{\rm Peak 2}$} &
No. & Peak Date & I$_{\rm median}$ & I$_{\rm peak}$ &{\bf a$_{\rm Peak 1}$} & {\bf m$_{\rm Peak 1}$}  &  {\bf a$_{\rm Peak 2}$} & {\bf m$_{\rm Peak 2}$} \\
\hline
1 & 01/17/2010 04:05 & -0.68 & -0.15 & -0.03 & -1.08 & ... & ... & 63 & 03/05/2012 23:00 & -0.76 & 0.42 & 1.87 & -1.49 & 2.27 & -1.77 \\
2 & 03/13/2010 14:00 & -1.22 & -0.76 & -0.47 & -1.41 & ... & ... & 64 & 03/05/2012 23:00 & -0.76 & 0.42 & 1.87 & -1.49 & 2.27 & -1.77 \\
3 & 06/12/2010 08:10 & -0.77 & 0.05 & -0.14 & -1.07 & ... & ... & 65 & 03/07/2012 15:35 & -0.76 & 3.19 & 3.71 & -0.59 & 5.96 & -2.12 \\
4 & 08/02/2010 09:15 & -0.78 & -0.39 & 1.12 & -1.46 & 0.43 & -0.99 & 66 & 03/09/2012 05:10 & -0.76 & 2.69 & 3.45 & -0.83 & 6.04 & -2.58 \\
5 & 08/08/2010 04:30 & -0.78 & -0.37 & 0.77 & -1.13 & -0.08 & -0.55 & 67 & 03/10/2012 18:55 & -0.76 & 1.87 & 2.59 & -0.81 & 6.48 & -3.44 \\
6 & 08/18/2010 12:25 & -0.78 & 0.34 & 1.45 & -1.39 & 1.95 & -1.73 & 68 & 03/13/2012 23:25 & -0.76 & 1.97 & 2.68 & -0.89 & 5.75 & -2.97 \\
7 & 08/31/2010 22:20 & -0.78 & -0.41 & -0.18 & -1.12 & ... & ... & 69 & 03/18/2012 23:35 & -0.76 & -0.37 & -0.46 & -0.81 & ... & ... \\
8 & 09/09/2010 03:40 & -0.79 & -0.19 & 0.53 & -0.81 & 0.61 & -0.87 & 70 & 03/21/2012 17:40 & -0.76 & -0.43 & -0.31 & -1.08 & ... & ... \\
9 & 01/13/2011 21:40 & -0.80 & -0.46 & -0.92 & -0.67 & ... & ... & 71 & 03/24/2012 16:25 & -0.76 & -0.30 & 1.03 & -1.38 & 0.36 & -0.93 \\
10 & 01/28/2011 09:20 & -0.80 & -0.02 & 0.87 & -0.89 & 1.55 & -1.35 & 72 & 03/25/2012 13:35 & -0.76 & -0.19 & 0.34 & -0.63 & 0.99 & -1.08 \\
11 & 01/28/2011 16:20 & -0.80 & 0.36 & 1.68 & -1.39 & 1.72 & -1.42 & 73 & 03/27/2012 18:25 & -0.76 & -0.21 & 1.06 & -1.39 & 0.95 & -1.32 \\
12 & 01/28/2011 16:20 & -0.80 & 0.36 & 1.68 & -1.39 & 1.72 & -1.42 & 74 & 03/28/2012 13:35 & -0.76 & -0.16 & -0.04 & -1.19 & ... & ... \\
13 & 01/31/2011 20:20 & -0.80 & -0.39 & -0.90 & -0.59 & ... & ... & 75 & 03/28/2012 13:35 & -0.76 & -0.16 & -0.04 & -1.19 & ... & ... \\
14 & 02/14/2011 03:30 & -0.81 & -0.42 & -0.72 & -0.63 & ... & ... & 76 & 04/08/2012 00:35 & -0.74 & -0.23 & -0.26 & -0.81 & ... & ... \\
15 & 02/15/2011 11:05 & -0.81 & 0.35 & 0.86 & -0.70 & 2.85 & -2.04 & 77 & 04/10/2012 03:55 & -0.74 & -0.16 & 1.24 & -1.47 & 0.20 & -0.77 \\
16 & 02/24/2011 21:30 & -0.81 & -0.30 & 0.85 & -1.22 & 0.90 & -1.26 & 78 & 04/15/2012 11:05 & -0.74 & -0.29 & 1.34 & -1.60 & -0.35 & -0.45 \\
17 & 03/08/2011 07:25 & -0.81 & 1.52 & 2.37 & -1.10 & 7.60 & -4.63 & 79 & 04/17/2012 09:35 & -0.74 & -0.26 & 0.53 & -0.87 & 1.29 & -1.38 \\
18 & 03/08/2011 07:25 & -0.81 & 1.52 & 2.37 & -1.10 & 7.60 & -4.63 & 80 & 04/19/2012 02:35 & -0.74 & -0.28 & 0.68 & -0.98 & 0.77 & -1.04 \\
19 & 03/22/2011 01:30 & -0.81 & 1.11 & 2.10 & -1.16 & 4.45 & -2.75 & 81 & 04/28/2012 03:50 & -0.74 & -0.35 & -0.12 & -1.22 & ... & ... \\
20 & 05/10/2011 10:30 & -0.84 & -0.36 & 0.28 & -0.81 & 1.21 & -1.44 & 82 & 05/07/2012 02:25 & -0.72 & -0.27 & 0.51 & -0.77 & 2.08 & -1.83 \\
21 & 05/30/2011 08:20 & -0.84 & -0.34 & -0.34 & -0.91 & ... & ... & 83 & 05/17/2012 04:35 & -0.72 & 2.49 & 2.81 & -0.48 & 5.41 & -2.24 \\
22 & 05/30/2011 08:20 & -0.84 & -0.34 & -0.34 & -0.91 & ... & ... & 84 & 05/27/2012 10:25 & -0.72 & 1.09 & 0.37 & -1.93 & ... & ... \\
23 & 06/02/2011 13:45 & -0.81 & -0.36 & -0.49 & -0.70 & ... & ... & 85 & 06/09/2012 10:40 & -0.78 & -0.33 & 0.82 & -1.24 & 0.73 & -1.18 \\
24 & 06/05/2011 00:50 & -0.81 & 0.50 & 2.36 & -1.84 & 3.22 & -2.43 & 86 & 06/10/2012 00:50 & -0.78 & -0.37 & 1.33 & -1.68 & 0.32 & -1.00 \\
25 & 06/05/2011 00:50 & -0.81 & 0.50 & 2.36 & -1.84 & 3.22 & -2.43 & 87 & 07/03/2012 01:05 & -0.61 & -0.14 & -0.11 & -0.97 & ... & ... \\
26 & 06/07/2011 18:20 & -0.81 & 1.83 & 2.60 & -0.84 & 4.35 & -2.02 & 88 & 07/05/2012 06:55 & -0.61 & -0.23 & -0.08 & -1.26 & ... & ... \\
27 & 06/07/2011 19:10 & -0.81 & 1.78 & 2.50 & -0.80 & 4.37 & -2.06 & 89 & 07/06/2012 00:10 & -0.61 & -0.24 & -0.11 & -1.06 & ... & ... \\
28 & 06/13/2011 11:40 & -0.81 & 0.62 & 0.21 & -1.66 & ... & ... & 90 & 07/07/2012 07:30 & -0.61 & 1.32 & 2.03 & -0.86 & 5.08 & -2.93 \\
29 & 07/26/2011 19:55 & -0.86 & -0.45 & 0.52 & -1.10 & 0.48 & -1.07 & 91 & 07/09/2012 04:20 & -0.61 & 1.25 & 2.02 & -0.86 & 4.21 & -2.34 \\
30 & 08/02/2011 11:40 & -0.82 & 0.43 & 1.13 & -0.86 & 2.82 & -2.01 & 92 & 07/12/2012 22:25 & -0.61 & 1.88 & 2.64 & -1.25 & 8.30 & -5.08 \\
31 & 08/04/2011 10:25 & -0.82 & 1.92 & 2.30 & -0.57 & 5.76 & -2.91 & 93 & 07/18/2012 06:00 & -0.61 & 2.02 & 2.89 & -1.37 & 6.92 & -4.10 \\
32 & 08/09/2011 09:15 & -0.82 & 1.47 & 1.79 & -0.39 & 3.78 & -1.73 & 94 & 07/18/2012 06:40 & -0.61 & 1.95 & 2.95 & -1.47 & 8.02 & -4.91 \\
33 & 08/09/2011 09:15 & -0.82 & 1.47 & 1.79 & -0.39 & 3.78 & -1.73 & 95 & 07/19/2012 15:00 & -0.61 & 1.89 & 2.63 & -0.90 & 6.67 & -3.63 \\
34 & 09/06/2011 13:45 & -0.82 & 0.40 & 0.97 & -0.76 & 4.13 & -2.90 & 96 & 07/23/2012 22:10 & -0.61 & 1.07 & 2.05 & -0.91 & 3.11 & -1.62 \\
35 & 09/07/2011 07:10 & -0.82 & 0.90 & 1.53 & -0.75 & 3.62 & -2.16 & 97 & 08/13/2012 01:15 & -0.83 & -0.29 & 0.72 & -1.19 & 1.05 & -1.41 \\
36 & 09/07/2011 19:00 & -0.82 & 0.54 & 1.28 & -0.88 & 2.89 & -1.97 & 98 & 08/22/2012 07:45 & -0.83 & -0.34 & 0.79 & -1.20 & 0.16 & -0.77 \\
37 & 09/08/2011 23:45 & -0.82 & -0.05 & 1.05 & -1.29 & 1.20 & -1.38 & 99 & 08/22/2012 07:45 & -0.83 & -0.34 & 0.79 & -1.20 & 0.16 & -0.77 \\
38 & 09/11/2011 00:40 & -0.82 & -0.43 & 0.87 & -1.31 & 0.52 & -1.08 & 100 & 08/31/2012 21:30 & -0.83 & -0.50 & -0.41 & -1.03 & ... & ... \\
39 & 09/23/2011 07:25 & -0.82 & 0.31 & 0.77 & -0.63 & 3.00 & -2.14 & 101 & 09/08/2012 16:30 & -0.81 & 0.16 & 1.14 & -1.00 & 3.53 & -2.62 \\
40 & 09/23/2011 07:25 & -0.82 & 0.31 & 0.77 & -0.63 & 3.00 & -2.14 & 102 & 09/20/2012 09:45 & -0.81 & -0.33 & 0.00 & -0.45 & 1.67 & -1.58 \\
41 & 09/25/2011 12:50 & -0.82 & 0.81 & 1.91 & -1.28 & 4.75 & -3.21 & 103 & 09/21/2012 08:15 & -0.81 & -0.35 & -0.17 & -1.19 & ... & ... \\
42 & 09/25/2011 12:55 & -0.82 & 0.86 & 2.04 & -1.53 & 3.40 & -2.45 & 104 & 09/28/2012 05:10 & -0.81 & 0.31 & 1.15 & -0.85 & 2.47 & -1.74 \\
43 & 09/25/2011 16:55 & -0.82 & 1.08 & 2.62 & -1.75 & 4.92 & -3.31 & 105 & 09/28/2012 05:10 & -0.81 & 0.31 & 1.15 & -0.85 & 2.47 & -1.74 \\
44 & 09/26/2011 02:45 & -0.82 & 1.35 & 3.05 & -1.99 & 5.40 & -3.58 & 106 & 09/29/2012 06:55 & -0.81 & -0.08 & 1.26 & -1.43 & 0.47 & -0.90 \\
45 & 09/30/2011 01:10 & -0.82 & -0.47 & -0.24 & -1.10 & ... & ... & 107 & 10/14/2012 17:50 & -0.87 & -0.41 & -0.44 & -0.78 & ... & ... \\
46 & 10/02/2011 08:25 & -0.85 & -0.30 & 0.74 & -1.05 & 0.82 & -1.11 & 108 & 10/22/2012 08:30 & -0.87 & -0.37 & -0.11 & -1.25 & ... & ... \\
47 & 10/22/2011 13:00 & -0.85 & -0.47 & 0.75 & -1.25 & 0.17 & -0.85 & 109 & 11/24/2012 08:35 & -0.87 & -0.33 & 0.90 & -1.26 & 0.88 & -1.25 \\
48 & 10/23/2011 09:55 & -0.85 & 0.33 & -0.03 & -1.30 & ... & ... & 110 & 12/05/2012 15:05 & -0.86 & -0.39 & -0.49 & -0.80 & ... & ... \\
49 & 11/04/2011 11:45 & -0.83 & 0.41 & 0.96 & -0.67 & 4.53 & -3.09 & 111 & 02/27/2013 08:20 & -0.83 & -0.10 & 1.00 & -1.20 & 1.74 & -1.70 \\
50 & 11/10/2011 05:15 & -0.83 & -0.42 & 0.54 & -1.01 & -0.15 & -0.54 & 112 & 03/05/2013 20:45 & -0.86 & -0.34 & -0.52 & -0.71 & ... & ... \\
51 & 11/27/2011 07:15 & -0.83 & 1.41 & 2.80 & -1.74 & 6.05 & -3.93 & 113 & 03/07/2013 13:30 & -0.86 & -0.26 & 1.16 & -1.50 & 0.73 & -1.21 \\
52 & 12/21/2011 23:40 & -0.84 & -0.44 & -0.32 & -0.99 & ... & ... & 114 & 03/16/2013 06:55 & -0.86 & 0.54 & 1.92 & -1.54 & 2.85 & -2.16 \\
53 & 12/24/2011 17:45 & -0.84 & -0.43 & -0.34 & -0.85 & ... & ... & 115 & 03/23/2013 21:30 & -0.86 & -0.36 & -0.43 & -0.88 & ... & ... \\
54 & 12/25/2011 21:55 & -0.84 & -0.30 & -0.28 & -0.85 & ... & ... & 116 & 03/23/2013 21:30 & -0.86 & -0.36 & -0.43 & -0.88 & ... & ... \\
55 & 01/02/2012 23:55 & -0.82 & -0.15 & 0.02 & -1.15 & ... & ... & 117 & 04/11/2013 16:50 & -0.86 & 2.02 & 2.61 & -0.78 & 5.36 & -2.64 \\
56 & 01/20/2012 09:45 & -0.82 & -0.33 & 0.73 & -0.96 & 0.63 & -0.90 & 118 & 04/18/2013 19:05 & -0.86 & -0.43 & -0.50 & -0.83 & ... & ... \\
57 & 01/24/2012 01:40 & -0.82 & 3.32 & 4.01 & -0.94 & 11.26 & -5.84 & 119 & 05/02/2013 01:25 & -0.87 & -0.32 & 0.78 & -1.19 & 0.50 & -1.01 \\
58 & 01/28/2012 12:15 & -0.82 & 2.36 & 2.79 & -0.57 & 6.63 & -3.18 & 120 & 05/14/2013 02:00 & -0.87 & -0.22 & -0.06 & -1.20 & ... & ... \\
59 & 01/28/2012 12:15 & -0.82 & 2.36 & 2.79 & -0.57 & 6.63 & -3.18 & 121 & 05/14/2013 13:00 & -0.87 & 0.14 & 1.48 & -1.35 & 2.68 & -2.17 \\
60 & 02/25/2012 09:40 & -0.83 & -0.18 & 1.55 & -1.63 & 1.04 & -1.28 & 122 & 05/16/2013 04:15 & -0.87 & 1.41 & 2.80 & -1.68 & 5.78 & -3.70 \\
61 & 03/05/2012 10:45 & -0.76 & 0.06 & 1.31 & -1.34 & 1.30 & -1.34 & 123 & 05/23/2013 06:50 & -0.87 & 3.22 & 3.94 & -1.13 & 8.76 & -4.39 \\
62 & 03/05/2012 10:45 & -0.76 & 0.06 & 1.31 & -1.34 & 1.30 & -1.34 & \\
\hline
\end{longtable*}

\begin{center}
\scriptsize The measured parameters from GOES-13 include the time of the peak proton flux intensity at $> 10$\,MeV within 24 hrs after the type II burst start time, the logarithm of the $> 10$\,MeV median flux during the month where the flux was below 1 pfu (I$_{\rm median}$ in pfu), the logarithm of the peak flux at $> 10$\,MeV (I$_{\rm peak}$ in pfu), and parameters of the best-fit power law or broken power law model to the energy spectrum at the peak proton flux (including the logarithm of the constant, {\bf a}, and the slope, {\bf m}, for points below $> 30$\,MeV (Peak1) and higher energies (Peak2).  For the first two bursts, GOES-11 observations were used since no data are available from GOES-15.
\end{center}

\clearpage
\begin{table}
\begin{center}
\caption{\small Summary of Properties of SEP/non-SEP Events.}\label{table-summary}
\begin{tabular}{l c c}
\hline\hline
{\bf Property} & {\bf SEP} & {\bf non-SEP} \\
\hline
Type II Peak Intensity (sfu) & $8.9 \times 10^2$ ($7.0 \times 10^4$)& $4.9 \times 10^2$ ($4.1 \times 10^4$)\\
Type II Integral Intensity (sfu) & $3.7 \times 10^3$ ($6.9 \times 10^4$)& $7.2 \times 10^3$ ($3.5 \times 10^5$)\\
\hline
Percent with Type III burst & 92\% & 59\% \\
Type III Peak Intensity (sfu) & $1.2 \times 10^7$ ($2.9 \times 10^7$)& $7.6 \times 10^5$ ($7.9 \times 10^6$)\\
Type III Duration (min) & 13.0 (11.8) & 3.0 (8.7)\\
\hline 
\end{tabular}
\end{center}

\begin{center}
\scriptsize The median and standard deviation, in parentheses, are given for the specified parameters for proton flux peaks $\ge 10$\,pfu (SEP) and $< 10$\,pfu (non-SEP).
\end{center}

\end{table}

\clearpage

\begin{table}
\begin{center}
\caption{\small Principal Component Results.}\label{table-pca}
\begin{tabular}{l c c}
\hline\hline
{\bf Variable} & {\bf 9 Variables} & {\bf 5 Variables} \\
\hline
Type II Peak Intensity & 0.346 & 0.444\\
Type II Integral Intensity & 0.288 & 0.373\\
Type II Duration & 0.206 & --\\
Type II Slope & 0.005 & --\\
Type II Frequency Range & 0.264 & --\\
\hline
Type III Intensity & 0.614 & 0.628\\
Type III Duration & 0.332 & 0.357\\
Type III Slope & 0.265 & --\\
\hline
Langmuir Peak Intensity & 0.356 & 0.376\\
\hline 
\end{tabular}
\end{center}

\begin{center}
\scriptsize The first component results from PCA of the radio burst parameters. The loading or relative weight on each of the nine or five feature-scaled input variables is shown. The scaled variables are multiplied by the weights/loading shown to yield the first component ($C1$), which accounts for the variance shown in the \% of Variance in C1 row. { These are new ``radio indices" that are used to predict whether an SEP event will occur.}
\end{center}
\end{table}

\begin{table}
\begin{center}
\caption{Logistic Regression Results. 
}\label{table-logistic}
\begin{tabular}{l c c c c c c}
\hline
{\bf Variables} & {\bf 9} & {\bf 9} & {\bf 5} & {\bf 5}  &  {SWPC} & Laurenza\\
{\bf Scaling} & {\bf Feature} & {\bf Log} &{\bf Feature} & {\bf Log} &  model & model \\
\hline
{\bf POD} & 62\% 	& 58\% 	& 58\% 	& 58\% 	& 54\% & 63\%\\
{\bf FAR} & 21\% 	& 22\% 	& 22\% 	& 22\% 	& 42\% & 42\%\\
{\bf PC} & 85\% 	& 84\% 	& 84\% 	& 84\% 	& -- & 93\%\\
{\bf HSS} & 0.60 	& 0.56 	& 	0.56	& 0.56  	& 0.55 & 0.58 \\
{\bf B$_0$} & -4.721 & -3.270 	& -4.053 	& -3.184 	& -- & --\\
{\bf B$_1$} & 3.327 	& 0.4227 	& 3.528 	& 0.422 	& -- & -- \\
\hline
\end{tabular}
\end{center}

\begin{center}
\scriptsize The statistics indicating probability of detection (POD), false alarm rate (FAR), percent correct (PC) and Heidke Skill Score (HSS) are shown for a logistic regression fit using the radio indices (C1; 9-variable feature-scaled and log-scaled and 5-variable feature-scaled and log-scaled)  to predict the probability of an SEP event.  The last two columns are the statistics from the SWPC Protons model and the {\it Laurenza et al.} (2009) model. Probability of an SEP event occurring, from the logistic regression models, is calculated as:
\begin{math} P({\rm SEP | C1}) = 1/(1 + e^{-({\rm B}_0 + {\rm B}_1 \rm{C1})}).\end{math} A probability $\ge 0.5$ predicts an SEP event. Otherwise, no SEP event is predicted.
\end{center}
\end{table}

\end{document}